\documentclass[final,1p,times]{elsarticle}

\usepackage{amsmath,amssymb,amsfonts,amscd}
\usepackage[mathscr]{eucal}

\newcommand{\set}[1]{\left\{ #1 \right\}}

\newcommand{\bbset}[1]{\mathbb{#1}}

\newcommand{\Integer}{\bbset{Z}}
\newcommand{\Real}{\bbset{R}}

\newcommand{\Complex}{\bbset{C}}

\newcommand{\pdfrac}[2]{\frac{\partial #1}{\partial #2}}

\newcommand{\pddiv}[2]{\partial #1 / \partial #2}

\newcommand{\Hilbert}{\mathcal{H}}

\newcommand{\bvec}[1]{\boldsymbol{#1}} % Bold Vector

\def\toexp{\mathop{\rm exp}}
\newcommand{\Texp}{\toexp_{\leftarrow}}
\newcommand{\Tprod}{{\prod_{\leftarrow}}}

\newcommand{\paulivec}{\bvec{\sigma}}

\newcommand{\bra}[1]{\langle{}#1{}|}
\newcommand{\ket}[1]{|{}#1{}\rangle}
\newcommand{\bracket}[2]{\langle{}#1{}|{}#2{}\rangle}

\newcommand{\IdH}{\hat{1}_{\Hilbert}}

\newcommand{\AF}{A}
\newcommand{\AD}{A^{\rm D}}
\newcommand{\AntiTexp}{\toexp_{\rightarrow}}

\newcommand{\Bmu}{B_{\mu}}
\newcommand{\Blambda}{B_{\lambda}}

\newcommand{\szp}{\uparrow}
\newcommand{\szm}{\downarrow}

\newcommand{\qi}{i}
\newcommand{\qj}{j}
\newcommand{\qk}{k}

\newcommand{\TR}{K}

\begin{document}

\begin{frontmatter}
  \title{A Unified Theory of Quantum Holonomies}

\author[label1]{Atushi Tanaka}
\ead{tanaka@phys.metro-u.ac.jp}
\address[label1]{Department of Physics, Tokyo Metropolitan University,
  Hachioji, Tokyo 192-0397, Japan}
\author[label2]{Taksu Cheon}
\ead{taksu.cheon@kochi-tech.ac.jp}
\address[label2]{Laboratory of Physics, Kochi University of Technology,
  Tosa Yamada, Kochi 782-8502, Japan}

\date{\today}

\begin{abstract}
  A periodic change of slow environmental parameters of a quantum
  system induces quantum holonomy.  The phase holonomy is a well-known
  example.  Another is a more exotic kind that exhibits eigenvalue and
  eigenspace holonomies.  We introduce a theoretical formulation that
  describes the phase and eigenspace holonomies on an equal footing.
  The key concept of the theory is a gauge connection for an ordered
  basis, which is conceptually distinct from Mead-Truhlar-Berry's
  connection and its Wilczek-Zee extension.  A gauge invariant treatment
  of eigenspace holonomy based on Fujikawa's formalism is developed.
  Example of adiabatic quantum holonomy, including the exotic kind with
  spectral degeneracy, are shown.
\end{abstract}

\begin{keyword}
%% keywords here, in the form: keyword \sep keyword
geometric phase \sep exotic hononomies \sep gauge theory 
%% PACS codes here, in the form: \PACS code \sep code
\PACS 03.65.Vf \sep 03.65.Ca \sep 42.50.Dv
%% MSC codes here, in the form: \MSC code \sep code
%% or \MSC[2008] code \sep code (2000 is the default)

\end{keyword}
%% PACS 2008
%% 03.65.Vf Phases: geometric; dynamic or topological
%% 03.65.Ca Formalism
%% 42.50.Dv Quantum state engineering and measurements (see also
%%          03.65.Ud Entanglement and quantum nonlocality, e.g.,
%%          EPR paradox, Bells inequalities, GHZ states, etc.)
%% cf. 03.67.Ac Quantum algorithms, protocols, and simulations

\end{frontmatter}

%%%
%%% Start of the main text
%%%

\section{Introduction}
\label{sec:introduction}

Consider a quantum system in a stationary state.
%TC
Let us adiabatically
change a parameter of the system along a closed path where
the spectral
%TC
degeneracy is assumed to be absent.
We ask the destination of 
the state after a change of the parameter along the path. 
This question is frequently raised in discussions 
of the Berry phase~\cite{Berry:PRSLA-392-45}. An answer, 
%which was supposed to be sensible one 
which is widely shared since Berry's work, is that 
a discrepancy remains in the phase of the state vector, 
even 
%TC
after the dynamical phase is excluded.
Indeed this is correct in a huge amount of 
examples~\cite{Shapere:GPP-1989,Bohm:GPQS-2003}. 
%and is called a {\em phase anholonomy}.
However, it is shown that this answer is not universal
in a recent report of exotic anholonomies~\cite{Cheon:PLA-248-285}
%TC
in which the initial and the final states are orthogonal in spite of
the absence of the spectral degeneracy.
In other words, the eigenspace associated with the adiabatic 
%TC
cyclic evolution exhibits discrepancy, or anholonomy.
Furthermore, the eigenspace discrepancy induces another discrepancy
in the corresponding eigenenergy. 

For the phase discrepancy, an established 
interpretation in terms of differential geometry allows us to call it 
the phase {\em holonomy}~\cite{Simon:PRL-51-2167}.
This interpretation naturally invites its non-Abelian extension,
which has been subsequently 
%TC
discovered by Wilczek and Zee in systems with spectral 
degeneracies~\cite{Wilczek:PRL-52-2111}.
Contrary to this, any successful association of the eigenspace 
%TC
discrepancy with the concept of holonomy has not been known.

The aim of this paper is to 
%TC
demonstrate that an interpretation of
the eigenspace discrepancy in terms of holonomy is indeed possible.
To achieve this, we introduce a framework that treats 
the phase and the eigenspace holonomies in a unified manner
in Section~\ref{sec:unifiedTheory}.
The key
%TC
concept is a non-Abelian gauge connection
that is associated with a parameterized basis~\cite{Filipp:PRA-68-02112},
and the identification of the place where the gauge connection
resides in the time evolution.
This is achieved through 
%TC
a fully gauge invariant extension of Fujikawa's formulation
that has been introduced for 
the phase holonomy~\cite{Fujikawa:AP-322-1500,Fujikawa:PRD-72-025009}.
Our approach is 
%TC
illustrated by the analysis of
adiabatic quantum holonomies of three examples.
First, Berry's Hamiltonian with spin-$\frac{1}{2}$ is revisited
in Section~\ref{sec:exampleBerry}.
The role of 
parallel transport~\cite{Stone:PRSLA-351-141,Simon:PRL-51-2167}, 
which accompanies the
multiple-valuedness of a parameterized basis, 
in our formulation will be emphasized.
The second example, shown in Section~\ref{sec:exampleAbelianCheon},
exhibits exotic holonomies without spectral degeneracy.
The last example, shown in Section~\ref{sec:exampleNonAbelianCheon},
is the simplest examples of the exotic holonomies
in the presence of degeneracy, 
i.e., the eigenspace holonomy {\'a} la Wilczek and Zee.
Section~\ref{sec:summary} provides a summary and an outlook.
A brief, partial report of the present result can be found in 
Ref.~\cite{Cheon:EPL-85-20001}.

\section{A gauge theory for a parameterized basis}
\label{sec:unifiedTheory}
%TC
Two building blocks of our theory, a gauge connection that is associated with a 
parameterized basis~\cite{Filipp:PRA-68-02112}, and 
Fujikawa formalism, originally conceived  for 
the phase holonomy, are presented in order to introduce our approach
to quantum holonomies.

\subsection{A gauge connection}
In the presence of the quantum holonomy, basis 
vectors are,
%TC
in general, multiple-valued as functions of a parameter.
In order to cope with such multiple-valuedness, we introduce 
a gauge connection for a parameterized basis.
This 
%TC
has been introduced by Filipp and Sj{\"o}qvist~\cite{Filipp:PRA-68-02112}
to examine 
Manini-Pistolesi off-diagonal geometric phase~\cite{Manini:PRL-85-3067}.
As is explained below, 
this gauge connection is different from 
Mead-Truhlar-Berry's~\cite{Mead:JCP-70-2284,Berry:PRSLA-392-45} and
Wilczek-Zee's gauge connections~\cite{Wilczek:PRL-52-2111},
which describe 
solely
%entirely 
the phase holonomy.

For $N$-dimensional Hilbert space $\Hilbert$, 
let $\set{\ket{\xi_n(s)}}_{n=0}^{N-1}$ be 
a complete orthogonal normalized system that is smoothly depends 
on a parameter $s$.
The parametric dependence induces
a gauge connection $\AF{}(s)$, which is
a $N\times{}N$ Hermite matrix and whose $(n,m)$-th element is
\begin{equation}
  \AF_{nm}(s)\equiv i\bra{\xi_{n}(s)}\pdfrac{}{s}\ket{\xi_{m}(s)}
  .
\end{equation}
By definition, $\AF{}(s)$ is non-Abelian.
For given $\AF(s)$, 
the basis vector $\ket{\xi_n(s)}$ obeys 
the following differential equation
\begin{align}
  \label{eq:EOMbyComponents}
  i\pdfrac{}{s}\ket{\xi_m(s)}&
  = \sum_n \AF_{nm}(s)\ket{\xi_n(s)} 
\end{align}
and we may solve this equation with an ``initial condition'' at $s=s'$.

The dynamical variable of the equation of motion~\eqref{eq:EOMbyComponents} is
an ordered sequence of basis vectors, also called a frame,
\begin{equation}
  \label{eq:fDef}
  f(s)
  \equiv
  \begin{bmatrix}
    \ket{\xi_0(s)},& \ket{\xi_1(s)},& \dots,& \ket{\xi_{N-1}(s)}
  \end{bmatrix}
  .
\end{equation}
Its conjugation 
\begin{equation}
  f(s)^{\dagger}
  =
  \begin{bmatrix}
    \bra{\xi_0(s)}\\ \bra{\xi_1(s)}\\ \dots\\ \bra{\xi_{N-1}(s)}
  \end{bmatrix}
\end{equation}
is also useful. 
For example, the resolution of unity by
$\set{\ket{\xi_n(s)}}_{n=0}^{N-1}$ is expressed as
\begin{equation}
  f(s)\left\{f(s)^{\dagger}\right\} = \IdH
  ,
\end{equation}
where $\IdH$ is the identical operator for $\Hilbert$,
and the gauge connection $\AF(s)$ is written as
\begin{align}
  \label{eq:defAF}
  \AF(s) 
  = i \left\{f(s)^{\dagger}\right\}\pdfrac{}{s}f(s)
  .
\end{align}
Now we have the equation of motion for $f(s)$
\begin{align}
  i\pdfrac{}{s} f(s)
  = f(s) \AF(s)
  .
\end{align}
Its formal solution is
\begin{align}
  f({s''})
  = f({s'})
  \AntiTexp\left(-i\int_{s'}^{s''} \AF(s)ds\right)
  ,
\end{align}
where $\AntiTexp$ is the anti-ordered exponential 
for the contour integration by $s$~\cite{Filipp:PRA-68-02112}.
Note that we need to specify the integration path to deal with 
the multiple-valuedness of $f(s)$, in general.

Our designation ``gauge connection'' for the whole $\AF(s)$ is
intended to clarify the difference from two famous gauge connections
for the phase holonomy.  When we choose
$\set{\ket{\xi_n(s)}}_{n=0}^{N-1}$ as an adiabatic basis of a
non-degenerate Hamiltonian $\hat{H}(s)$ with an adiabatic parameter
$s$, the elements of the gauge connection $\AF(s)$ have a well-known
interpretation: A diagonal element $\AF_{nn}(s)$ is
Mead-Truhlar-Berry's Abelian gauge connection for a single adiabatic
state $\ket{\xi_n(s)}$.  The off-diagonal elements are nonadiabatic
transition matrix elements, and, constitute ``the field strength''
corresponding to Mead-Truhlar-Berry's Abelian gauge
connection~\cite{Berry:PRSLA-392-45}.  This also applies to a
degenerate Hamiltonian with Wilczek-Zee's non-Abelian gauge
connection, which describes the change in an eigenspace.  On the other
hand, $\AF(s)$ that contains all elements $\set{\AF_{mn}(s)}$ is
defined with respect to the change of the frame $f(s)$, instead of
each eigenspace~\cite{Filipp:PRA-68-02112}.  As to be shown below,
$\AF(s)$ plays the central role in the quantum holonomy.

\subsection{An extended Fujikawa formalism}
Fujikawa has introduced a formulation 
to examine the quantum holonomy accompanying 
time evolution that involves a change of 
a parameter~\cite{Fujikawa:AP-322-1500,Fujikawa:PRD-72-025009}.
We will focus on the unitary time evolution for pure state in the following.
As a building block of the time evolution, 
we examine a parameterized quantum map, 
whose stroboscopic, unit time evolution 
from $\ket{\psi'}$ to $\ket{\psi''}$
is described by
\begin{equation}
  \label{eq:quantumMap}
  \ket{\psi''} = \hat{U}(s)\ket{\psi'}
  ,
\end{equation}
where $\hat{U}(s)$ is a unitary operator with a parameter $s$.
This is because periodically driven systems,
whose Floquet operator is $\hat{U}(s)$,
are our primary examples,
and our approach is immediately applicable to a Hamiltonian time evolution,
where $\hat{U}(s)$ correspond to an infinitesimal time evolution operator.
Let us vary $s$ along a path $C$ in the parameter space
during $L$ iterations of the quantum map~\eqref{eq:quantumMap},
where $s'$ and $s''$ is the initial and finial points, respectively.
Accordingly we examine the whole time evolution operator
\begin{align}
  \hat{U}(\set{s_l}_{l=0}^{L-1})
  \equiv\hat{U}(s_{L-1})\hat{U}(s_{L-2})\cdots{}\hat{U}(s_{0})
  ,
\end{align}
where $s_l$ is the value of $s$ at $l$-th step.
Although the present formulation is applicable to investigate
nonadiabatic settings, our primary interest 
here is an adiabatic behavior induced by the limiting
procedure $L\to\infty$.
The ``$f(s)$-representation''  of 
the building block $\hat{U}(s)$ of the whole evolution is
a $N\times{}N$ unitary matrix
\begin{align}
  Z(s)
  \equiv\left\{f(s)^{\dagger}\right\}\hat{U}(s) f(s)
  .
\end{align}
In other words, we have
$\hat{U}(s)= f(s) Z(s)\left\{f(s)^{\dagger}\right\}$.
In order to deal with the change of $s$ from $s_l$ to $s_{l+1}$, we have
\begin{align}
  \hat{U}(s_l) &
  = f(s_{l+1})\left\{f(s_{l+1})^{\dagger}\right\}\times\hat{U}(s_l)
  \nonumber
  \\ &
  = 
  f(s_{l+1}) Z_{\rm F}(s_{l+1}, s_{l})
  \left\{f(s_l)^{\dagger}\right\}
  ,
\end{align}
where an effective time evolution matrix 
$Z_{\rm F}(s_{l+1}, s_{l})$ incorporates the unit dynamical
evolution and the parametric change of 
$s$ along a part of path $C$
\begin{align}
  Z_{\rm F}(s_{l+1}, s_{l})
  \equiv 
  %\Torder\left[
    \Texp\left(i\int_{s_l}^{s_{l+1}} \AF(s)ds\right)
  %\right]
  Z(s_l)
  ,
\end{align}
where $\Texp$ is the ordered exponential.
Hence the whole time evolution is expressed as
\begin{align}
  \hat{U}(\set{s_l}_{l=0}^{L-1})
  = f(s'')B_{\rm d}(\set{s_l}_{l=0}^{L})\left\{f(s')^{\dagger}\right\}
\end{align}
where we have an ``effective'' time evolution operator
\begin{align}
  \label{eq:defEffEvol}
  B_{\rm d}(\set{s_l}_{l=0}^{L})&
  \equiv
  Z_{\rm F}(s_{L}, s_{L-1})
  Z_{\rm F}(s_{L-1}, s_{L-2})
  \dots
  %\nonumber{}\\{}&\qquad{}\times
  Z_{\rm F}(s_{1}, s_{0})
  .
\end{align}
Finally, we have ``the $f(s')$-representation'' of 
the whole time evolution operator
\begin{align}
  \label{eq:fRepresentationOfUwhole}
  \hat{U}(\set{s_l}_{l=0}^{L-1})
  = f(s') W(C) B_{\rm d}(\set{s_l}_{l=0}^{L})
  \left\{f(s')^{\dagger}\right\}
  ,
\end{align}
where
\begin{align}
  W(C)&
  \equiv\AntiTexp\left(-i\int_{C} \AF(s)ds\right)
  .
\end{align}

Since the above definitions are exact, our formulation is invariant
against a basis transformation with 
$N\times{}N$ unitary matrix $G(s)$
\begin{align}
  \label{eq:fGaugeTrans}
  f(s)\mapsto{}f(s)G(s)
\end{align}
once we incorporate the following transformations
\begin{subequations}
\begin{align}
  \AF(s)&
  \mapsto 
  G(s)^{\dagger} \AF(s) G(s) + i G(s)^{\dagger} \pdfrac{G(s)}{s}
  ,\\
  %Z(s) &
  %\mapsto G(s)^{\dagger} Z(s) G(s)
  %,\\
  W(C)&\mapsto
  G(s')^{\dagger}W(C)G(s'')
  ,\\
  B_{\rm d}(\set{s_l}_{l=0}^{L})&\mapsto
  G(s'')^{\dagger}B_{\rm d}(\set{s_l}_{l=0}^{L})G(s')
  .
\end{align}
\end{subequations}
This is Fujikawa's 
hidden local gauge invariance~\cite{Fujikawa:PRD-72-025009} 
in a generalized form.
The strategy of Fujikawa formalism is to extract a geometric information
from the whole time evolution operator 
via the expression~\eqref{eq:fRepresentationOfUwhole} with
an appropriate restriction of $G(s)$, as is shown below.

Let us examine the case that the change of $s$ is slow enough 
so that we may employ the adiabatic approximation~\cite{Born:ZP-51-165}.
Accordingly it is suitable to choose $f(s)$ as an adiabatic basis, 
i.e., each basis vector $\ket{\xi_n(s)}$ is 
an eigenvector of $\hat{U}(s)$, 
to make $Z(s)$ a diagonal matrix, 
whose non-zero elements are the eigenvalues of $\hat{U}(s)$.
Let $z_n(s)$ be the eigenvalue of $\hat{U}(s)$ corresponding
to an eigenvector $\ket{\xi_n(s)}$,
i.e.,
\begin{equation}
  \hat{U}(s)\ket{\xi_n(s)} = z_n(s)\ket{\xi_n(s)}
  ,
\end{equation}
where we assume that there is no degeneracy in eigenvalue.
Note that $z_n(s)$ is unimodular due to the unitarity of $\hat{U}(s)$.
Now the gauge transformation $G(s)$ is restricted to 
$U(1)^{\otimes N}$ times a permutation matrix, which correspond
to the freedoms to choose the phases of basis vectors, and
assign the quantum numbers, respectively. The permutation matrix is 
required to deal with the eigenspace holonomy, as is shown below.
In terms of $Z_{\rm F}(s_{l+1}, s_{l})$, the adiabatic approximation
is the diagonal approximation~\cite{Fujikawa:PRD-72-025009}
\begin{align}
  &
  Z_{\rm F}(s_{l+1}, s_{l})
  %\nonumber\\{} &
  \simeq 
  Z^{\rm D}_{\rm F}(s_{l+1}, s_{l})
  \equiv \Texp\left(i\int_{s_l}^{s_{l+1}} \AD(s)ds\right) Z(s_l)
  ,
\end{align}
where $\AD(s)$ is the diagonal part of the gauge connection $\AF(s)$,
i.e., $\AD_{mn}(s)= \delta_{mn}\AF_{mm}(s)$.
Namely, $\AD_{mm}(s)$ is Mead-Truhlar-Berry's gauge connection
for the $m$-th state 
$\ket{\xi_m(s)}$~\cite{Mead:JCP-70-2284,Berry:PRSLA-392-45}.
The corresponding adiabatic approximation of
$B_{\rm d}(\set{s_l}_{l=0}^{L})$~\eqref{eq:defEffEvol} is
\begin{align}
  B_{\rm ad}(\set{s_l}_{l=0}^{L})&
  \equiv
  Z^{\rm D}_{\rm F}(s_{L}, s_{L-1}) Z^{\rm D}_{\rm F}(s_{L-1}, s_{L-2})\dots
  %\nonumber\\{}&\qquad{}\times
  Z^{\rm D}_{\rm F}(s_{1}, s_{0})
  ,
\end{align}
which are decomposed into two parts:
\begin{subequations}
\begin{align}
  B_{\rm ad}(\set{s_l}_{l=0}^{L})&
  = B(C) D(\set{s_l}_{l=0}^{L})
  ,
  \intertext{where}
  B(C)&
  \equiv\Texp\left(i\int_{C} \AD(s)ds\right)
\intertext{is the geometric part, and,}
  D(\set{s_l}_{l=0}^{L})&
  \equiv\prod_{l=0}^{L-1}Z(s_l)
\end{align}
\end{subequations}
contains dynamical phases.
We retain the path-ordered exponential in $B(C)$ to make it applicable
to the cases with the presence of spectral degeneracies, as shown below.
To sum up, the adiabatic approximation of the whole time evolution is
\begin{align}
  \label{eq:adiabaticTimeEvolutionOperator}
  \hat{U}(\set{s_l}_{l=0}^{L-1})
  \simeq f({s'}) M(C) D(\set{s_l}_{l=0}^{L}) \left\{f(s')^{\dagger}\right\}
  ,
\end{align}
where
\begin{align}
  \label{eq:defM}
  M(C)&
  \equiv W(C)B(C)
  \nonumber\\{}&
  = \AntiTexp\left(-i\int_{C} \AF(s)ds\right)
  \Texp\left(i\int_{C} \AD(s)ds\right)
  ,
\end{align}
is the geometric part determined by the path $C$, and
two gauge connections $\AF(s)$ and $\AD(s)$.

For a cyclic path $C$, we call $M(C)$~(Eq.~\eqref{eq:defM})
{\em a holonomy matrix}, which describes the adiabatic change of state
vector, starting from an eigenstate at $s=s'$, along the 
closed path $C$.

An explanation why $W(C)$ is required to describe the eigenspace holonomy
is the following.
Let us assume that $f(s)$ is single-valued.
This implies that $W(C)$ is the $N\times{}N$ identical matrix.
Consequently, $M(C) = B(C)$ is always diagonal and thus cannot 
describe the eigenspace holonomy,
though the single-valuedness assumption does not prevent
the conventional approach from describing the phase holonomy.
On the other hand, in the presence of the eigenspace holonomy,
a factor of $M(C)$ need to be a permutation matrix.
Since $B(C)$, which is always a diagonal matrix according to its definition,
cannot be such a factor, 
the permutation matrix is need to be supplied by $W(C)$.
This is consistent with the fact that the presence of 
the eigenspace anholonomy implies the multiple-valuedness of $f(s)$.

Furthermore, when we employ the parallel transport condition
$\AD(s)=0$~\cite{Stone:PRSLA-351-141,Simon:PRL-51-2167,Filipp:PRA-68-02112}, 
the holonomy matrix takes extremely simple form:
\begin{align}
  \label{eq:MinPT}
  M^{\rm p.t.}(C)&
  =\AntiTexp\left(-i\int_{C} \AF(s)ds\right)
  ,
\end{align}
which is determined only by $W(C)$. In other words, 
all the adiabatic quantum holonomies can be summarized as 
a holonomy in the orderd basis $f(s)$~(Eq.~\eqref{eq:fDef}).
For the phase holonomy, this observation is already reported 
by Fujikawa~\cite{Fujikawa:AP-322-1500,Fujikawa:PRD-72-025009}.
In this sense, the parallel transport condition offers
a privileged gauge.

We explain the consequence of the gauge 
transformation~\eqref{eq:fGaugeTrans} under the adiabatic approximation,
where $G(s)$ is restricted to be a product of
a permutation matrix and a diagonal unitary matrix.
The invariance of the adiabatic time evolution operator, which appears
at the right hand side of Eq.~\eqref{eq:adiabaticTimeEvolutionOperator},
is assured due to the following
\begin{subequations}
\begin{align}
  \AD(s)&
  \mapsto G(s)^{\dagger} \AD(s) G(s) + i G(s)^{\dagger} \pdfrac{G(s)}{s}
  ,\\
  W(C)&
  \mapsto G(s')^{\dagger} W(C) G(s'')
  ,\\
  B(C)&
  \mapsto G(s'')^{\dagger} B(C) G(s')
  .
\end{align}
\end{subequations}
Hence we obtain the manifest covariance of $M(C)$
\begin{align}
  M(C)&
  \mapsto G(s')^{\dagger} M(C) G(s')
  .
\end{align}

Eq.~\eqref{eq:defM} provides a correct expression 
of the holonomy matrix $M(C)$
not only for maps~\eqref{eq:quantumMap} but also for
flows, i.e., Hamiltonian systems and periodically driven systems.
For a system whose time evolution is generated by
a nearly static Hamiltonian $\hat{H}(s)$, we will derive Eq.~\eqref{eq:defM}
in Appendix~\ref{app:MforHamiltonian} with a suitable discretization of time.
For a periodically driven system described by a Hamiltonian 
$\hat{H}(t, s)$,
where $\hat{H}(t, s)=\hat{H}(t+T,s)$ is assumed, a Floquet 
operator
\begin{align}
  \Texp\left(-\frac{i}{\hbar}\int_0^{T} \hat{H}(t,s) dt\right)
\end{align}
is the unitary operator $\hat{U}(s)$ to provide a stroboscopic 
description of the system. Hence this system is reduced to a quantum map.

An extension of our formulation to the case that the presence of spectrum 
degeneracy whose degree is independent with $s$
along a closed path $C$ is shown.
The resultant expression for the holonomy matrix~\eqref{eq:defM} 
remains the same.
This is achieved by a suitable extension of gauge connections
$\AF(s)$ and $\AD(s)$.
For the eigenspace corresponding to the eigenvalue $z_n(s)$ of $\hat{U}(s)$,
we have a normalized orthogonal vectors $\ket{\xi_{n\nu}(s)}$
with an index $\nu$ for the eigenspace,
where
\begin{align}
  \hat{U}(s)\ket{\xi_{n\nu}(s)} = z_n(s)\ket{\xi_{n\nu}(s)}
\end{align}
and 
$
\bracket{\xi_{n''\nu''}(s)}{\xi_{n'\nu'}(s)} 
= \delta_{n''n'}\delta_{\nu''\nu'}
$.
The gauge connection for the parameterized basis is
\begin{align}
  \AF_{n''\nu'', n'\nu'}(s)&
  \equiv i\bra{\xi_{n''\nu''}(s)}\pdfrac{}{s}\ket{\xi_{n'\nu'}(s)}
  .
\end{align}
It is straightforward to see that 
$\AD(s)$ that appears in Eq.~\eqref{eq:defM} 
is non-Abelian:
\begin{align}
  \AD_{n''\nu'', n'\nu'}(s)&
  \equiv \delta_{n''n'} \AF_{n''\nu'', n'\nu'}(s)
  ,
\end{align}
where $\AD_{n\nu'', n\nu'}(s)$ is Wilcek-Zee's gauge connection
for the $n$-th eigenspace~\cite{Wilczek:PRL-52-2111}.

\section{An example of the phase holonomy}
\label{sec:exampleBerry}

Our formulation is applied to Berry's simplest example of the
adiabatic phase holonomy~\cite{Berry:PRSLA-392-45}. 
Let us suppose that a spin-$\frac{1}{2}$ is 
under static magnetic field $\bvec{B}$.
With a suitable choice of units,
the spin is described by a Hamiltonian 
\begin{equation}
  \label{eq:HBerry}
  \hat{H}(\bvec{B}) 
  = \bvec{B}\cdot\hat{\paulivec}
  ,
\end{equation}
where
$\hat{\bvec\sigma} = \sum_{j=x,y,z}\hat{\sigma}_j\bvec{e}_j$ is 
the Pauli operator for the spin, and,
$\bvec{e}_j$ ($j=x,y,z$) is the unit vector for $j$-axis.
The spectrum of $\hat{H}(\bvec{B})$ is
$\set{\pm B}$, where $B\equiv\|\bvec{B}\|$.
To investigate the adiabatic holonomy, we have to exclude the degeneracy point
$B=0$.
The unit vector $\bvec{n}\equiv\bvec{B}/B$ is parameterized with
the spherical coordinate, i.e.,
$\bvec{n}
= \bvec{e}_x\cos\varphi\sin\theta +
\bvec{e}_y\sin\varphi\sin\theta + \bvec{e}_z\cos\theta$.
Let $\ket{\xi_{\pm}(\bvec{B})}$ be a normalized eigenvector
of $\hat{H}(\bvec{B})$, corresponding to the eigenvalue $\pm B$:
\begin{subequations}
\begin{align}
  \ket{\xi_+(\bvec{B})}&
  = e^{-i\varphi/2}\cos\frac{\theta}{2}\ket{\szp}
  + e^{i\varphi/2}\sin\frac{\theta}{2}\ket{\szm}
  \\
  \ket{\xi_-(\bvec{B})}&
  = -e^{-i\varphi/2}\sin\frac{\theta}{2}\ket{\szp}
  + e^{i\varphi/2}\cos\frac{\theta}{2}\ket{\szm}
  .
\end{align}
\end{subequations}
Note that $\ket{\xi_{\pm}(\bvec{B})}$ are multiple-valued 
as functions of $(\theta,\varphi)$. To assure them single-valued,
the range of $(\theta,\varphi)$ needs to be restricted
within an open set
$U_{\rm M}
\equiv \set{(\theta,\varphi)\vert
  \text{$0 < \theta < \pi$ and $0 < \varphi < 2\pi$}}$, for example.
The corresponding frame 
$f(\bvec{B}) 
  \equiv
  \begin{bmatrix}
    \ket{\xi_+(\bvec{B})},& \ket{\xi_-(\bvec{B})}
  \end{bmatrix}
$
is
\begin{align}
  \label{eq:fBerry}
  f(\bvec{B}) &
  =
  \begin{bmatrix}
    \ket{\szp},& \ket{\szm}
  \end{bmatrix}
  \begin{bmatrix}
    e^{-i\varphi/2}\cos\frac{\theta}{2}&
    -e^{-i\varphi/2}\sin\frac{\theta}{2}
    \\
    e^{i\varphi/2}\sin\frac{\theta}{2}&
    e^{i\varphi/2}\cos\frac{\theta}{2}
  \end{bmatrix}
  .
\end{align}
Gauge connections 
$A_x(\bvec{B}) \equiv 
i\left\{f(\bvec{B})^{\dagger}\right\}\pddiv{f(\bvec{B})}{x}$
($x = \theta, \varphi, B$)
for parametric changes of 
$\bvec{B}$ are
\begin{subequations}
\begin{align}
  \AF_{\theta}(\bvec{B})&
  %\equiv i\left\{f(\bvec{B})^{\dagger}\right\}\pdfrac{f(\bvec{B})}{\theta}
  = \frac{1}{2}\sigma_y
  ,
  \\
  \label{eq:BAFphi}
  \AF_{\varphi}(\bvec{B})&
  %\equiv i\left\{f(\bvec{B})^{\dagger}\right\}\pdfrac{f(\bvec{B})}{\varphi}
  = \frac{1}{2}\left(\sigma_z \cos\theta - \sigma_x\sin\theta\right)
  ,
  \\
  \AF_{B}(\bvec{B})&
  %\equiv i\left\{f(\bvec{B})^{\dagger}\right\}\pdfrac{f(\bvec{B})}{B}
  = 0
  ,
\end{align}
\end{subequations}
where we employ $2\times2$ complex matrices
\begin{gather}
  \label{eq:def2time2matrices}
  %I_2
  %\equiv\begin{bmatrix}1&0\\{}0&1\end{bmatrix}
  %,\quad
  \sigma_x
  \equiv\begin{bmatrix}0&1\\{}1&0\end{bmatrix}
  ,\quad
  \sigma_y
  \equiv\begin{bmatrix}0&-i\\{}i&0\end{bmatrix}
  ,\quad
  \sigma_z
  \equiv\begin{bmatrix}1&0\\{}0&-1\end{bmatrix}
  .
\end{gather}
Accordingly, the Mead-Truhlar-Berry gauge connections are
\begin{align}
  \label{eq:BAD}
  \AD_{\theta}(\bvec{B}) = 0
  ,\quad
  \AD_{\varphi}(\bvec{B}) = \frac{1}{2}\sigma_z\cos\theta
  ,\quad
  \AD_{B}(\bvec{B}) = 0
  .
\end{align}
As is well known,
the strength of the magnetic field
$B$ plays no particular role in the computation of the phase 
holonomy once $B$ is kept nonzero.

Evaluations of the holonomy matrix $M(C)$ 
for typical closed loops in the parameter space are shown.
First, we examine a loop $C$ in which $f(\bvec{B})$ is single-valued,
e.g. $C\subset U_{\rm M}$.
Consequently $W(C)$ is the $2\times2$ identical matrix.
This is a conventional wisdom to obtain a formula of the phase
holonomy, where an appropriate gauge for $f(\bvec{B})$ 
(or, equivalently, $\ket{\xi_{\pm}(\bvec{B})}$)
against the loop is chosen to avoid the multiple-valuedness of 
$f(\bvec{B})$~\cite{Berry:PRSLA-392-45}.
Hence all the holonomies reside in $B(C) = M(C)$.
The evaluation of the contour integral in $B(C)$ is straightforward
to obtain the classic result in the matrix $M(C)$:
\begin{align}
  \label{eq:MBerry}
  M(C)&
  = \exp\left(-\frac{i}{2} \Omega(C) \sigma_z\right)
  ,
\end{align}
where $\Omega(C)$ is the solid angle for $C$~\cite{Berry:PRSLA-392-45}.

Second, we examine a meridian great circle $C_{\theta}$ in which
$\theta$ moves $0$ to $2\pi$ with $\varphi$ kept fixed.
The parametric change along such a circle can induce a change of
the sign of an eigenvector~\cite{Herzberg:DFS-35-77}.
In our example, $f(\bvec{B})$ cannot be single-valued on $C_{\theta}$:
\begin{align}
  \left.f(\bvec{B})\right|_{\theta\downarrow0}
  = - \left.f(\bvec{B})\right|_{\theta\uparrow2\pi}
  .
\end{align}
Although the conventional strategy mentioned above is to 
avoid such multi-valuedness,
we insist the present choice of the gauge~\eqref{eq:fBerry}
to show an alternative way to reproduce the conventional result.
Thanks to the present choice of the gauge, the Mead-Truhlar-Berry
gauge connection satisfies the parallel transport condition
$\AD_{\theta} =0$ and $B(C_{\theta}) = 1$.
This enable us to employ Eq.~(\ref{eq:MinPT}) to obtain the holonomy matrix:
\begin{align}
  M(C_{\theta}) &
  = \AntiTexp\left(-i\oint_{C_{\theta}} \frac{1}{2}\sigma_y d\theta\right)
  = e^{-i\pi\sigma_y}
  = -1
  ,
\end{align}
which is consistent with Eq.~\eqref{eq:MBerry}.
This is an example that our formulation properly deals
the multiple-valuedness of $f(\bvec{B})$.
On the other hand, if we choose an appropriate gauge to
make $f(\bvec{B})$ single-valued on the circuit $C_{\theta}$, 
$W(C_{\theta})$ is trivial to put all the nontrivial holonomy in 
$B(C_{\theta})$, as is stated above.

Finally, let us consider a circle of latitude $C_{\varphi}$,
where $\theta$ is kept fixed and $\varphi$ is increased from $0$ to $2\pi$.
It is straightforward to obtain
$B(C_{\varphi}) = \exp(i\pi\sigma_z\cos\theta)$
and $W(C_{\varphi}) = -1$.
The latter indicates again the sign change in the parametric dependence
along $C_{\varphi}$.
These elements are combined to reproduce a well-known result
\begin{align}
  M(C_{\varphi}) = \exp\left\{-i\pi(1-\cos\theta)\sigma_z\right\}
  .
\end{align}
We remark that, in all the examples above,
the holonomy matrices $M$ are diagonal,
so that the eigenspace holonomy is absent.
Accordingly the eigenvalue anholonomy is also absent.

\section{An example of the exotic holonomies
  in quantum map spin-$\frac{1}{2}$}
\label{sec:exampleAbelianCheon}

In Berry's Hamiltonian (Eq.~(\ref{eq:HBerry})), 
the strength of the magnetic field $B$ plays no role 
in quantum holonomies. 
One reason is that the corresponding gauge connection $\AF_{B}(\bvec{B})$ 
vanishes. Another reason is that it is impossible to make any loop
in the parameter space by an increment of $B$, 
with $\bvec{n}$ being kept fixed.
To make a loop 
for a strength parameter,
%against the strength, 
we may examine the following 
quantum map for a spin-$\frac{1}{2}$
\begin{align}
  \label{eq:UPrototype}
  \exp\left\{-i\lambda\left(a + b\bvec{n}\cdot\hat{\bvec\sigma}\right)\right\}
  ,
\end{align}
where $\bvec{n}$ is a normalized real vector,
$a$ and $b$ are real constants 
to be
%being 
specified later,
and $\lambda$ is the strength.
The periodicity of the quantum map 
with respect to
%against 
the increment of $\lambda$
implies that there is a loop in the parameter space of $\lambda$.
In particular, if we choose $a = q/2$ and $b= (2-q/2)$ with 
an integer $q$, Eq.~\eqref{eq:UPrototype} is periodic 
as a function of $\lambda$, with a primitive period $2\pi$.
Accordingly, the parameter space of $\lambda$ is identified with
$S^{1}$ and it might be suitable to investigate quantum holonomies 
for a periodic variation of $\lambda$.
However, such a loop does not allow us to study adiabatic holonomies,
since there remains a spectral degeneracy along the loop
at $\lambda = 0\; (\text{mod $2\pi$})$.

A simple way to lift the degeneracy at $\lambda=0$ is 
to concatenate two quantum maps:
\begin{align}
  \label{eq:asymmetricQuantumMap1/2}
  &
  \exp\left\{-i\mu\left(\frac{q}{2} 
      + \frac{2-q}{2}\bvec{m}\cdot\hat{\bvec\sigma}\right)\right\}
  %\nonumber\\{}&\quad{}\times
  \exp\left\{-i\lambda\left(\frac{p}{2} 
      + \frac{2-p}{2}\bvec{n}\cdot\hat{\bvec\sigma}\right)\right\}
  ,
\end{align}
where $q$ and $p$ are integers, 
$\bvec{m}$ and $\bvec{n}$ are normalized vectors in $\Real^3$,
and, $\mu$ and $\lambda$ are strengths.
Due to the periodicity in $\mu$ and $\lambda$, the parameter space
of $(\mu, \lambda)$ is a two-dimensional torus $S^1\times S^1$.
Both $\bvec{m}$ and $\bvec{n}$ specify points on a sphere $S^2$.
In the following, we fix $\bvec{m}=\bvec{e}_z$ and 
parameterize $\bvec{n}$ by spherical variables $\theta$ and $\varphi$
as $\bvec{n}
= \bvec{e}_x\cos\varphi\sin\theta +
  \bvec{e}_y\sin\varphi\sin\theta + \bvec{e}_z\cos\theta$.
If we change $\bvec{m}$ with keeping $\bvec{m}\cdot\bvec{n}$ fixed,
it induces only the Berry phase.

To facilitate the following analysis, we examine the symmetric version of
the quantum map~\eqref{eq:asymmetricQuantumMap1/2}
\begin{align}
  \label{eq:quantumMapSpin1/2}
  \hat{U}&
  \equiv
  \exp\left\{-i\frac{\mu}{2}\left(\frac{q}{2} 
      + \frac{2-q}{2}\bvec{m}\cdot\hat{\bvec\sigma}\right)\right\}
  %\nonumber\\{}&\qquad{}\times
  \exp\left\{-i\lambda\left(\frac{p}{2} 
      + \frac{2-p}{2}\bvec{n}\cdot\hat{\bvec\sigma}\right)\right\}
  \nonumber\\{}&\qquad{}\times
  \exp\left\{-i\frac{\mu}{2}\left(\frac{q}{2} 
      + \frac{2-q}{2}\bvec{m}\cdot\hat{\bvec\sigma}\right)\right\}
  .
\end{align}
For brevity, we omit the parameters in the following.
A possible implementation of the quantum map~\eqref{eq:quantumMapSpin1/2}
is available by a periodically driven system that is described by 
the following Hamiltonian
\begin{align}
  \hat{H}(t)&
  \equiv
  \mu\left(\frac{q}{2} + \frac{2-q}{2}\bvec{m}\cdot\hat{\bvec\sigma}\right)
  %\nonumber\\{}&\qquad
  +
  \lambda\left(\frac{p}{2}+\frac{2-p}{2}\bvec{n}\cdot\hat{\bvec\sigma}\right)
  \sum_{j\in\Integer} \delta(t-j)
  ,
\end{align}
where the Floquet operator for a unit time interval 
$-1/2 \le t < 1/2$ is $\hat{U}$.
The magnitudes of the magnetic fields of the unperturbed system 
and the perturbation are
\begin{equation}
  \label{eq:defBmu}
\begin{aligned}
  \Bmu&\equiv \frac{1}{2}{(2-q)}\mu,\\
  \Blambda&\equiv \frac{1}{2}{(2-p)}\lambda,
\end{aligned}
\end{equation}
respectively.
It is straightforward to show 
\begin{align}
  \hat{U}&
  \equiv
  e^{-i(\mu q +\lambda p)/2}
  \left(\cos\frac{\Delta}{2} - i\hat{\bvec\sigma}\cdot\bvec{\tilde{l}}\right)
  ,
\end{align}
where
\begin{subequations}
\begin{align}
  \label{eq:DeltaQuantumMap1/2}
  \Delta
  &
  \equiv2\cos^{-1}
  \left(\cos\Bmu\cos\Blambda -\cos\theta\sin\Bmu\sin\Blambda\right)
  ,
  \\
  \bvec{\tilde{l}}
  &
  \equiv
  \left(\sin\Bmu\cos\Blambda+\cos\theta\cos\Bmu\sin\Blambda\right)
  \bvec{m}
  \nonumber\\ &\qquad
  {}+(\sin\Blambda)
  \left\{\bvec{n} - (\bvec{n}\cdot\bvec{m})\bvec{m}\right\}
  .
\end{align}
\end{subequations}
It is also easy to see $\|\bvec{\tilde{l}}\|^2 = \sin^2(\Delta/2)$.
Hence $\bvec{l}\equiv \bvec{\tilde{l}}/\sin(\Delta/2)$ is a unit vector.
Now we have
\begin{align}
  \hat{U}&
  \equiv\exp\left\{-i\left(\frac{\mu q +\lambda p}{2}
      +\frac{\Delta}{2}\hat{\bvec\sigma}\cdot\bvec{l}\right)\right\}
  ,
\end{align}
and its eigenvalues are
\begin{align}
  z_{\pm}\equiv 
  \exp\left\{-i\left(\frac{\mu q +\lambda p}{2}
      \pm \frac{\Delta}{2}\right)\right\}
  .
\end{align}
Corresponding quasienergies are
$E_{\pm} \equiv ({\mu q +\lambda p}\pm{\Delta})/{2}$,
which is defined up to modulus $2\pi$.

In order to study the adiabatic holonomies,
we need to identify the spectral degeneracies, whose condition 
is $e^{i\Delta} = 1$, in the parameter space.
It is useful to see $\Delta$ as a function of 
$\Blambda$
\begin{align}
  \label{eq:DeltaVsNu}
  \Delta = 2\cos^{-1}
  \left(A \cos\left(\Blambda + \tilde{a}\right) \right)
\end{align}
where 
$A\equiv \sqrt{1 - \sin^2\theta\sin^2\Bmu}$
and $\tilde{a}$ is an ``initial phase'' that is independent with $\lambda$.
We choose the branch of $\cos^{-1}A$ in $[0,\pi]$.
If $A < 1$,
$\Delta$ oscillates within the range 
$[2\cos^{-1}A, 2\pi-2\cos^{-1}A] \subset (0, 2\pi)$
as a function of $\Blambda$, and encounters no spectral degeneracy.
On the other hand, the condition
$A=1$ (i.e., $\sin\theta\sin\Bmu=0$) implies
the presence of the spectral degeneracy.
From the similar argument for $\Bmu$, we will encounter 
spectral degeneracies if $\sin\theta\sin\Bmu\sin\Blambda=0$.

Let us examine the case $\sin\theta=0$.
Since this implies $\cos\theta=\pm 1$,
we have
\begin{align}
  \Delta 
  = 2\cos^{-1}\left(\cos\left(\Blambda\pm\Bmu\right)\right)
  .
\end{align}
Accordingly the degeneracy points draw
lines in 
$(\Bmu, \Blambda)$-plane as
\begin{align}
  \label{eq:dgeneracyLatticeLinesBerry}
  \set{(\Bmu, \Blambda) \Bigm|
    \frac{\Blambda\pm\Bmu}{\pi}\in\Integer}
\end{align}
corresponding to the condition $\cos\theta=\pm 1$.
On the other hand, if we assume
$\sin\Bmu\sin\Blambda = 0$, we have
\begin{align}
  \Delta
  =2\cos^{-1}\left(\cos\Bmu\cos\Blambda\right)
  .
\end{align}
Hence another condition for the spectral degeneracy is
$|\cos\Bmu\cos\Blambda| = 1$,
i.e., the degeneracy points are 
at lattice points:
\begin{align}
  \label{eq:dgeneracyLatticePtsBerry}
  \set{\left(\Bmu, \Blambda\right)
    \Bigm| \frac{\Bmu}{\pi}, \frac{\Blambda}{\pi}\in \Integer}%
  ,
\end{align}
for all $\theta$.

Summarizing above, we show the location of the spectral degeneracies
in terms of $(\mu,\nu)$, whose space is the two-dimensional torus.
The degeneracy lines are
specified by $(\mu, \lambda, \theta)$ as
\begin{align}
  &
  \set{(\mu, \lambda, \theta)\Bigm|
    \frac{\lambda(2-p)+\mu(2-q)\cos\theta}{2\pi}\in\Integer,
    \frac{\theta}{\pi}\in\Integer}
  .
\end{align}
In addition to this, we have isoleted degeneracy points as
\begin{align}
  \set{\left(\mu=\frac{2\pi k}{2-q}, \lambda=\frac{2\pi l}{2-p}\right)\Bigm|
    k,l\in \Integer}
  .
\end{align}

Except these degeneracy points, it is legitimate to introduce
a zenith angle $\Theta$ of $\bvec{l}$, s.t., 
\begin{equation}
  \label{eq:CosThetaQuantumMap1/2}
\begin{aligned}
  \cos\Theta &
  = \frac{\sin\Bmu\cos\Blambda+\cos\theta\cos\Bmu\sin\Blambda}%
  {\sin(\Delta/2)}
  ,
  \\
  %\label{eq:SinThetaQuantumMap1/2}
  \sin\Theta &
  = \frac{\sin\theta\sin\Blambda}{\sin(\Delta/2)}
  .
\end{aligned}
\end{equation}
%%
% \begin{subequations}
% \begin{align}
%   \label{eq:CosThetaQuantumMap1/2}
%   \cos\Theta &
%   = \frac{1}{\sin(\Delta/2)}
%   \left(\sin\frac{\mu(2-q)}{2}\cos\frac{\lambda(2-p)}{2}
%     +\cos\theta\cos\frac{\mu(2-q)}{2}\sin\frac{\lambda(2-p)}{2}\right)
%   \\
%   \label{eq:SinThetaQuantumMap1/2}
%   \sin\Theta &
%   = \frac{1}{\sin(\Delta/2)}\sin\theta\sin\frac{\lambda(2-p)}{2}
%   .
% \end{align}
% \end{subequations}
% From our choice of $\bvec{m}$ and $\bvec{n}$, 
% we have
% \begin{align}
%   \bvec{l}
%   = \bvec{e}_x\cos\varphi\sin\Theta +
%   \bvec{e}_y\sin\varphi\sin\Theta + \bvec{e}_z\cos\Theta
%   .
% \end{align}
It is straightforward to obtain 
the eigenvectors $\ket{\xi_{\pm}}$ of $\hat{U}$,
corresponding to the eigenvalues $z_{\pm}$:
\begin{subequations}
\begin{align}
  \ket{\xi_+}&
  = e^{-i\varphi/2}\cos\frac{\Theta}{2}\ket{\szp}
  + e^{i\varphi/2}\sin\frac{\Theta}{2}\ket{\szm}
  ,
  \\
  \ket{\xi_-}&
  = -e^{-i\varphi/2}\sin\frac{\Theta}{2}\ket{\szp}
  + e^{i\varphi/2}\cos\frac{\Theta}{2}\ket{\szm}
  .
\end{align}
\end{subequations}
Let 
$f \equiv 
\begin{bmatrix}
  \ket{\xi_+},&\ket{\xi_-}
\end{bmatrix}
$ be a frame.
The gauge connections~\eqref{eq:defAF} for $f$
are
\begin{subequations}
\begin{align}
  \AF_{\theta}&
  %\equiv if^{\dagger}\pdfrac{f}{\Theta}\pdfrac{\Theta}{\theta}
  = \frac{1}{2}\sigma_y\pdfrac{\Theta}{\theta}
  ,
  \\
  \AF_{\varphi}&
  %\equiv if^{\dagger}\pdfrac{f}{\varphi}
  = \frac{1}{2}\left(\sigma_z \cos\Theta - \sigma_x\sin\Theta\right)
  ,
  \\
  \AF_{\lambda}&
  %\equiv if^{\dagger}\pdfrac{f}{\Theta}\pdfrac{\Theta}{\lambda}
  = \frac{1}{2}\sigma_y\pdfrac{\Theta}{\lambda}
  ,
  \\
  \AF_{\mu}&
  %\equiv if^{\dagger}\pdfrac{f}{\Theta}\pdfrac{\Theta}{\mu}
  = \frac{1}{2}\sigma_y\pdfrac{\Theta}{\mu}
  ,
\end{align}
\end{subequations}
and the corresponding Mead-Truhlar-Berry gauge connections are 
$\AD_{\theta} = 0$,
$\AD_{\varphi}
= \frac{1}{2}\sigma_z \cos\Theta$,
$\AD_{\lambda} = 0$,
and 
$\AD_{\mu} = 0$.
With these gauge connections,
we will examine the holonomy matrices of typical loops in the parameter space.

First, we examine the meridian great circle $C_{\theta}$ in which
$\theta$ moves $0$ to $2\pi$ with other parameters are kept fixed.
It is straightforward to see $B(C_{\theta}) = 1$,
due to the parallel transport condition $\AD_{\theta} = 0$. 
Hence all the holonomies reside in $W(C_{\theta}) = M(C_{\theta})$:
\begin{align}
  \label{eq:WofCMinQuamtumMap1/2}
  W(C_{\theta}) &
  = \AntiTexp\left(-i\oint_{C_{\theta}}
    \frac{1}{2}\sigma_y\pdfrac{\Theta}{\theta}\;
    d\theta\right)
  \nonumber
  \\ &
  = \exp
  \left(-i\frac{1}{2}\sigma_y\left.\Theta\right|_{\theta=0}^{2\pi}\right)
  ,
\end{align}
where $\left.\Theta\right|_{\theta=0}^{2\pi}$,
the change of $\Theta$ along $C_{\theta}$, is determined by
the image of $C_{\theta}$ in the sphere $(\Theta, \varphi)$.
If the image is a closed circle, we have 
$\left.\Theta\right|_{\theta=0}^{2\pi}=\pm 2\pi$, where $\pm$
correspond to the direction of the path. 
Both cases provides a 
Longuet-Higgins type phase change 
$M(C_{\theta})= e^{\mp i\pi\sigma_y}=-1$.
On the other hand, if the image is closed self-retracing curve 
along an ark, we have 
$\left.\Theta\right|_{\theta=0}^{2\pi}=0$, 
which implies $M(C_{\theta})=1$.
The following index
\begin{align}
  \label{eq:defindexr}
  r %% was p
  \equiv 
  \left[\frac{\Blambda+\Bmu}{\pi}\right] 
  - \left[\frac{\Blambda-\Bmu}{\pi}\right],
  %%\left[\frac{\lambda(2-p)+\mu(2-q)}{2\pi}\right] 
  %%- \left[\frac{\lambda(2-p)-\mu(2-q)}{2\pi}\right],
\end{align}
where $[x]$ is the maximum integer not greater than $x$,
determines which is the case, as shown in 
Appendix~\ref{app:meridianKickedTop1/2}:
\begin{align}
  \label{eq:ThetaOfCMinQuamtumMap1/2}
  \left.\Theta\right|_{\theta=0}^{2\pi}
  =
  \begin{cases}
    2\pi (-1)^{r/2}& \text{for $r$ is even}\\
    0& \text{for $r$ is odd}
  \end{cases}
  .
\end{align}
Hence we obtain
\begin{align}
  \label{eq:MofCMinQuamtumMap1/2}
  M(C_{\theta})
  =
  (-1)^{1+r}
%   =
%   \exp\left\{i\pi
%     \left(1 + \left[\frac{\lambda(2-p)+\mu(2-q)}{2\pi}\right]
%       -\left[\frac{\lambda(2-p)-\mu(2-q)}{2\pi}\right]\right)\right\}
  .
\end{align}

Next, we examine a circle of latitude $C_{\varphi}$,
where  $\varphi$ is increased from $0$ to $2\pi$ and the other
parameters are kept fixed.
We have
$B(C_{\varphi}) = \exp(i\pi\sigma_z\cos\Theta)$
and $W(C_{\varphi}) = -1$.
Accordingly, we obtain
\begin{align}
  M(C_{\varphi}) = \exp\left(-i\pi(1-\cos\Theta)\sigma_z\right)
  .
\end{align}
So far, the holonomy matrices $M(C)$ are diagonal, 
so that
neither $C_{\theta}$ nor $C_{\varphi}$ incorporates the exotic holonomies.
The following is the first example of the exotic holonomies in this paper.

Let us examine a closed loop $C_{\lambda}$, in which
$\lambda$ is increased from $0$ to $2\pi$, being kept fixed other parameters.
To avoid degeneracies along $C_{\lambda}$, 
we choose 
$\mu (2-q)/(2\pi)\notin\Integer$ 
and $\theta/\pi\notin\Integer$.
%% This is valid only for $p=1$
%From Eq.~\eqref{eq:DeltaVsNu},
%the primitive period of $\Delta$ as a function of $\lambda$ is 
%$4\pi/|2-p|$, which is larger than $2\pi$ if $p\ne 0,4$.
When we increase $\lambda$ from $\lambda=\lambda'$ to $\lambda=\lambda'+2\pi$, 
we have 
\begin{align}
  \left.\Delta\right|_{\lambda=\lambda'+2\pi}
  = 
  \begin{cases}
    \left.\Delta\right|_{\lambda=\lambda'}& \text{for even $p$}\\
    2\pi - \left.\Delta\right|_{\lambda=\lambda'}& \text{for odd $p$}
  \end{cases}
  ,
\end{align}
i.e., an anholonomy in $\Delta$ occurs.
Accordingly we have an eigenvalue holonomy
\begin{align}
  \left.z_{\pm}\right|_{\lambda=\lambda'+2\pi}
  =
  \begin{cases}
    \left.z_{\pm}\right|_{\lambda=\lambda'}& \text{for even $p$}\\
    \left.z_{\mp}\right|_{\lambda=\lambda'}& \text{for odd $p$}
  \end{cases}
  ,
\end{align}
which implies the presence of the eigenspace holonomy.
We proceed to evaluate the holonomy matrix $M(C_{\lambda})$.
Because of the parallel transport condition $\AD_{\lambda}=0$, 
we have $M(C_{\lambda})=W(C_{\lambda})$.
On the other hand,
we have
%\begin{subequations}
\begin{align}
  W(C_{\lambda})&
  =\exp\left(-i\oint_{C_{\lambda}} 
    \frac{1}{2}\sigma_y\pdfrac{\Theta}{\lambda} d\lambda\right)
  \nonumber\\ &
  =\exp
  \left(-i\frac{1}{2}\sigma_y\left.\Theta\right|_{\lambda=0}^{\lambda=2\pi}\right)
  .
\end{align}
%\end{subequations}
Hence we need to examine the zenith angle $\Theta$.
From
Eq.~\eqref{eq:CosThetaQuantumMap1/2}
and
$\left.\sin(\Delta/2)\right|_{\lambda=\lambda'+2\pi}
= \left.\sin(\Delta/2)\right|_{\lambda=\lambda'}$,
we have 
$\left.e^{i\Theta}\right|_{\lambda=2\pi}
= (-1)^{(2-p)}\left.e^{i\Theta}\right|_{\lambda=0}$, i.e.,
$\left.\Theta\right|_{\lambda=0}^{2\pi} = \pi(2-p) \mod 2\pi$.
Since this is not suffice to determine the precise value of
$W(C_{\lambda})$, we need keep track of $\Theta$ against the increment 
of $\lambda$ to obtain $\left.\Theta\right|_{\lambda=0}^{2\pi}$.
Now the parameter space perpendicular to $C_{\lambda}$ is
$(\theta, \varphi, \mu)\in S^2 \times S^1$, which is
divided into subspaces by the spectral degeneracies
$\mu = 2\pi k / (2-q)$ ($k$ is integer).
Since, in each subspace, $\left.\Theta\right|_{\lambda=0}^{2\pi}$ is 
constant, it is suffice to evaluate it at a representative point.
Let us choose a point $\theta=\pi/2$ and
$\mu = \pi (2k + 1)/(2-q)$,
where the spectral gap takes a constant value $\Delta = \pi$,
from Eq.~\eqref{eq:DeltaQuantumMap1/2}.
Accordingly we have 
$
  e^{i\Theta} 
  = (-1)^k \exp\left\{i (-1)^k \lambda(2-p)/2\right\}
  ,
$
from Eq.~\eqref{eq:CosThetaQuantumMap1/2}.
Hence we obtain
\begin{align}
  \left.\Theta\right|_{\lambda=0}^{2\pi}
  = (-1)^k \pi(2-p) 
  ,
\end{align}
which also holds for 
$
2\pi k / (2-q) < \mu < 2\pi (k+1) / (2-q)
,
$
i.e., $k=[\mu(2-q)/(2\pi)]$.
Hence we have
%\begin{subequations}
\begin{align}
  M(C_{\lambda})&
  =\exp\left(-i\frac{1}{2}(-1)^{k} \pi(2-p)
    \sigma_y\right)
  \nonumber\\ &
  = 
  \begin{bmatrix}
    -\cos\frac{p\pi}{2},& -(-1)^{k}\sin\frac{p\pi}{2}\\
    (-1)^{k}\sin\frac{p\pi}{2},& -\cos\frac{p\pi}{2}
  \end{bmatrix}
  .
\end{align}
%\end{subequations}
In particular, if $p$ is odd, the off-diagonal elements of $M(C_{\lambda})$
remains, so that the eigenspace holonomy exhibits.
This is consistent with the emergence of the eigenvalue holonomy
for odd $p$.

Now the similar analysis of quantum holonomies for the circuit
$C_{\mu}$, where $\mu$ is increased from $0$ to $2\pi$, is trivial.
Hence we show only the holonomy matrix
\begin{align}
  M(C_{\mu})&
  = 
  \begin{bmatrix}
    -\cos\frac{q\pi}{2},& -(-1)^{k}\sin\frac{q\pi}{2}\\
    (-1)^{k}\sin\frac{q\pi}{2},& -\cos\frac{q\pi}{2}
  \end{bmatrix}
  ,
\end{align}
where $k=[\lambda(2-p)/(2\pi)]$. 
We conclude that odd $q$ along $C_{\mu}$ implies the exotic holonomies.

\section{Example 3: the exotic holonomies a l\'a Wilczek-Zee}
\label{sec:exampleNonAbelianCheon}
A simple example of the eigenspace holonomy accompanying spectral 
degeneracy is shown. 
Extending Mead's study~\cite{Mead:PRL-59-161,Avron:CMP-124-595} 
on non-Abelian adiabatic phase holonomy~\cite{Wilczek:PRL-52-2111},
we introduce a quantum map with Kramer's degeneracy.

\subsection{Quantum map for spin-$\frac{3}{2}$ 
  with Kramers' degeneracy}
\label{subsec:quantumMap3/2}
To introduce our model, we review the time-reversal invariance structure
in an atom with 
odd-number of electrons~\cite{Mead:PRL-59-161,Avron:CMP-124-595}.
For a comprehensive explanation, we refer 
Avron {\em et al.}~\cite{Avron:CMP-124-595}.
Let $\bvec{\hat{J}}$ be the total angular momentum of our system
and $\ket{J,M}$ the standard basis vector for $\bvec{\hat{J}}$, 
i.e., $\bvec{\hat{J}}^2 \ket{J,M} = J(J+1)\ket{J,M}$,
$\bvec{\hat{J}}\cdot\bvec{e}_z \ket{J,M} = M\ket{J,M}$,
and 
$\bvec{\hat{J}}\cdot(\bvec{e}_x \pm i\bvec{e}_y)\ket{J,M}
= \sqrt{J(J+1) - M(M\pm 1)}\ket{J,M\pm 1}$.
The standard time-reversal operator for $\bvec{\hat{J}}$ is
an anti-unitary operator 
$\hat{\TR}\equiv \exp(-i\pi\hat{J}_y) \hat{K}_0$, where
$\hat{\TR}_0$ is the complex conjugate operation in
the $\ket{J,M}$-representation.
We examine the fermion case $\hat{K}^2 = -1$, which implies that
$J$ is a half-integer.
If an Hermite operator commutes with $\hat{\TR}$, its spectrum
exhibits Kramer's degeneracy. The same is true for unitary operators.

We focus on the case $J=\frac{3}{2}$, and introduce basis vectors
as follows:
\begin{equation}
  \label{eq:defBasisE}
  \begin{aligned}
  \ket{e_1}&\equiv \ket{\frac{3}{2},\frac{3}{2}},
  \quad&
  \ket{\TR e_1}&\equiv 
  \hat\TR (\ket{e_1}) = \ket{\frac{3}{2},-\frac{3}{2}},
  \\
  \ket{e_2}&\equiv \ket{\frac{3}{2},-\frac{1}{2}},
  \quad&
  \ket{\TR e_2}&\equiv 
  \hat\TR (\ket{e_2}) = \ket{\frac{3}{2},\frac{1}{2}}
  .
  \end{aligned}
\end{equation}
Our physical observables are spanned by the following time-reversal
invariant operators
\begin{equation}
  \label{eq:defTau}
  \begin{aligned}
  \hat{\tau}_0&
  \equiv\ket{e_1}\bra{e_1}+\ket{\TR e_1}\bra{\TR e_1}
  - \ket{e_2}\bra{e_2}-\ket{\TR e_2}\bra{\TR e_2}
  %\nonumber\\{}&\qquad{}
  %+ \ket{e_2}(-1)\bra{e_2}+\ket{\TR e_2}(-1)\bra{\TR e_2}
  ,\\
  \hat{\tau}_1&
  \equiv\ket{e_1}\bra{\TR e_2}+\ket{\TR e_1}(-1)\bra{e_2}
  +\text{h.c.}%
  %\nonumber\\{}&\qquad{}
  %+ \ket{e_2}(-1)\bra{\TR e_1}+\ket{\TR e_2}\bra{e_1}
  ,\\
  \hat{\tau}_2&
  \equiv\ket{e_1}(-i)\bra{\TR e_2}+\ket{\TR e_1}(-i)\bra{e_2}
  +\text{h.c.}%
  %\nonumber\\{}&\qquad{}
  %+ \ket{e_2}i\bra{\TR e_1}+\ket{\TR e_2}i\bra{e_1}
  ,\\
  \hat{\tau}_3&
  \equiv\ket{e_1}\bra{e_2}+\ket{\TR e_1}\bra{\TR e_2}
  +\text{h.c.}%
  %\nonumber\\{}&\qquad{}
  %+ \ket{e_2}\bra{e_1}+\ket{\TR e_2}\bra{\TR e_1}
  ,\\
  \hat{\tau}_4&
  \equiv\ket{e_1}(-i)\bra{e_2}+\ket{\TR e_1}i\bra{\TR e_2}
  +\text{h.c.}%
  %\nonumber\\{}&\qquad{}
  %+ \ket{e_2}i\bra{e_1}+\ket{\TR e_2}(-i)\bra{\TR e_1}
  ,
  \end{aligned}
\end{equation}
which are traceless and form a Clifford algebra
\begin{align}
  \label{eq:CliffordAlgebra}
  \hat\tau_{\alpha}\hat\tau_{\beta}
  + \hat\tau_{\beta}\hat\tau_{\alpha}
  = 2\delta_{\alpha\beta}
  .
\end{align}
Several properties of $\tau_{\alpha}$ are shown in 
Appendix~\ref{app:AlgebraicPropsTau}.

We introduce an extension of the quantum map for spin-$\frac{1}{2}$
(Eq.~\eqref{eq:quantumMapSpin1/2})
\begin{align}
  \label{eq:quantumMapSpin3/2}
  \hat{U}&
  \equiv
  \exp\left\{-i\frac{\mu}{2}\left(\frac{q}{2} 
      + \frac{2-q}{2}\hat\tau_0\right)\right\}
  %\nonumber\\ &\qquad{}\times
  \exp\left\{-i\lambda\left(\frac{p}{2} 
      + \frac{2-p}{2}\sum_{\alpha=0}^4
      n_{\alpha}\hat\tau_{\alpha}\right)\right\}
  \nonumber\\ &\qquad{}\times
  \exp\left\{-i\frac{\mu}{2}\left(\frac{q}{2} 
      + \frac{2-q}{2}\hat\tau_0\right)\right\}
  ,
\end{align}
where $(n_{\alpha})_{\alpha=0}^4$ is a unit vector
in $\Real^5$, i.e., $\sum_{\alpha} n_{\alpha}^2=1$,
and $(q, p)\in\Integer^2$.
The quantum map~\eqref{eq:quantumMapSpin3/2} can be implemented
by a periodically pulsed driven system in a similar way shown
in the previous section for the quantum map~\eqref{eq:quantumMapSpin1/2}.
Since the unitary operator~\eqref{eq:quantumMapSpin3/2} is
$2\pi$-periodic both in $\mu$ and $\lambda$, 
the parameter space of $(\mu, \lambda)$ forms a two-dimensional
torus $S^1\times S^1$.
The unit vector $(n_{\alpha})$ 
is parameterized by spherical variables:
\begin{gather}
  n_0=\cos\theta
  ,\quad
  \begin{bmatrix} n_1\\ n_2\end{bmatrix}
  = \begin{bmatrix}\cos\chi\\ \sin\chi\end{bmatrix}\sin\eta\sin\theta
  ,
  \quad
  %\nonumber\\
  \begin{bmatrix} n_3\\ n_4\end{bmatrix}
  = \begin{bmatrix}\cos\varphi\\ \sin\varphi\end{bmatrix}\cos\eta\sin\theta
  .
\end{gather}

In Appendix~\ref{app:QM3/2Tilde}, we show 
\begin{align}
  \label{eq:QuantumMap3/2Tilde}
  \hat{U}&
  = e^{-i (\mu q + \lambda p)/2}
  \left(\cos\frac{\Delta}{2}
    - i\sum_{\alpha}\tilde{l}_{\alpha}\hat\tau_{\alpha}\right)
  ,
\end{align}
where $\Delta$ is defined in Eq.~\eqref{eq:DeltaQuantumMap1/2},
\begin{subequations}
\begin{align}
  \label{eq:DeltaQuantumMap3/2}
  \tilde{l}_0&
  = \sin\Bmu\cos\Blambda
  +\cos\theta\cos\Bmu\sin\Blambda%
  %%= \sin\frac{\mu(2-q)}{2}\cos\frac{\lambda(2-p)}{2}
  %\nonumber\\{}&\qquad
  %+\cos\theta\cos\frac{\mu(2-q)}{2}\sin\frac{\lambda(2-p)}{2}
  ,
  \intertext{and}
  \tilde{l}_{\alpha}&
  \equiv 
  n_{\alpha}\sin\Blambda
  %%n_{\alpha}\sin\frac{\lambda(2-p)}{2}
\end{align}
%%
% \begin{subequations}
% \begin{align}
%   \label{eq:DeltaQuantumMap3/2}
%   \Delta&
%   \equiv 2 \cos^{-1}\left(\cos\frac{\mu(2-q)}{2}\cos\frac{\lambda(2-p)}{2}
%     - \cos\theta\sin\frac{\mu(2-q)}{2}\sin\frac{\lambda(2-p)}{2}\right)
%   ,\\
%   \tilde{l}_0&
%   = \sin\frac{\mu(2-q)}{2}\cos\frac{\lambda(2-p)}{2}
%   +\cos\theta\cos\frac{\mu(2-q)}{2}\sin\frac{\lambda(2-p)}{2}
%   ,
%   \intertext{and}
%   \tilde{l}_{\alpha}&
%   \equiv n_{\alpha}\sin\frac{\lambda(2-p)}{2}
% \end{align}
\end{subequations}
for $\alpha\ne 0$,
where the definitions of $\Bmu$ and $\Blambda$ are 
shown in Eq.~(\ref{eq:defBmu}).
Since $\sum_{\alpha=0}^4 \tilde{l}_{\alpha}^2 = \sin(\Delta/2)^2$,
we normalize $\tilde{l}_{\alpha}$:
\begin{align}
  l_{\alpha}\equiv \frac{1}{\sin(\Delta/2)}\tilde{l}_{\alpha}
  .
\end{align}
Accordingly we have
\begin{align}
  \hat{U}&
  = \exp\left\{-i\left(\frac{\mu q + \lambda p}{2}
      + \frac{\Delta}{2}\sum_{\alpha}{l}_{\alpha}\hat\tau_{\alpha}
    \right)\right\}
  .
\end{align}
The eigenvalues of $\hat{U}$ are
\begin{align}
  z_{\pm}
  \equiv 
  \exp\left\{-i\left(\frac{\mu q + \lambda p}{2} \pm \frac{\Delta}{2}
    \right)\right\}
  .
\end{align}
Namely, the spectrum is completely same with the example shown in
Section~\ref{sec:exampleAbelianCheon}.
%%
%AT-2009-02
To examine eigenvectors, we parametrize $l_{\alpha}$
with the zenith angle $\Theta$,
which is already introduced for the quantum map spin-$\frac{1}{2}$
in Eq.~\eqref{eq:CosThetaQuantumMap1/2}.
We have
\begin{gather}
  l_0 
  = \cos\Theta
  ,\quad
  \begin{bmatrix} l_1\\ l_2\end{bmatrix}
  = \begin{bmatrix}\cos\chi\\ \sin\chi\end{bmatrix}
  \sin\eta\sin\Theta
  ,
  \quad%\nonumber\\
  \begin{bmatrix} l_3\\ l_4\end{bmatrix}
  = \begin{bmatrix}\cos\varphi\\ \sin\varphi\end{bmatrix}
  \cos\eta\sin\Theta
  .
\end{gather}
From Appendix~\ref{app:DiagonalizationOfN}, 
the eigenvectors of $\hat{U}$ are
\begin{equation}
  \begin{aligned}
  \ket{\xi_+}&
  = \ket{d_1}\cos\frac{\Theta}{2} + \ket{d_2}\sin\frac{\Theta}{2},
  \\
  \ket{\TR \xi_+}&
  = \ket{\TR d_1}\cos\frac{\Theta}{2} + \ket{\TR d_2}\sin\frac{\Theta}{2},
  \\
  \ket{\xi_-}&
  = \ket{d_1}\left(-\sin\frac{\Theta}{2}\right) 
  + \ket{d_2}\cos\frac{\Theta}{2},
  \\
  \ket{\TR \xi_-}&
  = \ket{\TR d_1}\left(-\cos\frac{\Theta}{2}\right) 
  + \ket{\TR d_2}\cos\frac{\Theta}{2}
  ,
  \end{aligned}
\end{equation}
where
\begin{equation}
\begin{aligned}
  \ket{d_1}&
  \equiv \ket{e_1}\left(e^{-i(\varphi+\chi)/2}\cos\frac{\eta}{2}\right)
  %\\{}&\qquad
  +\ket{\TR e_1}\left(-e^{+i(\varphi+\chi)/2}\sin\frac{\eta}{2}\right)
  ,
  \\
  \ket{d_2}&
  \equiv
  \ket{e_2}\left(e^{i(\varphi-\chi)/2}\cos\frac{\eta}{2}\right)
  %\\{}&\qquad
  +\ket{\TR e_2}\left(e^{-i(\varphi-\chi)/2}\sin\frac{\eta}{2}\right)
  ,\\
  \ket{\TR d_1}&
  = \ket{e_1}\left(+e^{-i(\varphi+\chi)/2}\sin\frac{\eta}{2}\right)
  %\\{}&\qquad
  + \ket{\TR e_1}\left(e^{+i(\varphi+\chi)/2}\cos\frac{\eta}{2}\right)
  ,
  \\
  \ket{\TR d_2}&
  = 
  \ket{e_2}\left(-e^{+i(\varphi-\chi)/2}\sin\frac{\eta}{2}\right)
  %\\{}&\qquad
  +\ket{\TR e_2}\left(e^{-i(\varphi-\chi)/2}\cos\frac{\eta}{2}\right)
  .
\end{aligned}
\end{equation}
Note that we put each basis vector before its complex coefficient above
to prevent a confusion due to the presence of anti-Hermite operation $K$.
To conclude this subsection, we introduce a frame
composed by the eigenvectors $\hat{U}$~\cite{endnote:CEPL}:
% \endnote{%
%   We remark on the difference from the previous 
%   report~\cite{Cheon:EPL-85-20001} on the notations concerning 
%   to the analysis of the kicked spin-$\frac{3}{2}$.
%   The spherical variables 
%   $\gamma$, $\eta$, $\xi$ and $\zeta$ in Ref~\cite{Cheon:EPL-85-20001}
%   should be read as $\theta$, $\frac{\pi}{2}-\eta$, $\chi$ and $\varphi$, 
%   respectively.
%   %%
%   The set of normalized eigenvectors
%   $\ket{v_{mn}}$ ($m, n = 0,1$) in Eq.~(36) of
%   Ref.~\cite{Cheon:EPL-85-20001} and 
%   $\ket{\xi_{\pm}}$ and $\ket{\TR \xi_{\pm}}$ are related by a unitary
%   transformation:
%   $\ket{v_{00}} 
%   = \ket{\xi_+}\cos\frac{\eta}{2}+ \ket{\TR\xi_+}\sin\frac{\eta}{2}$,
%   $\ket{v_{01}} 
%   = \ket{\xi_+}\left(-\sin\frac{\eta}{2}\right)
%   + \ket{\TR\xi_+}\cos\frac{\eta}{2}$,
%   $\ket{v_{10}} 
%   = \ket{\xi_-}\cos\frac{\eta}{2}
%   + \ket{\TR\xi_-}\left(-\sin\frac{\eta}{2}\right)$,
%   and
%   $\ket{v_{11}} 
%   = \ket{\xi_-}\sin\frac{\eta}{2}
%   + \ket{\TR\xi_-}\cos\frac{\eta}{2}$.
%   Accordingly the choices of basis vectors have difference.
%   This modifies the expressions of the gauge connections and holonomy 
%   matrices.
% }:
\begin{equation}
  f\equiv 
  \begin{bmatrix}
    \ket{\xi_+},& \ket{K \xi_+},&\ket{\xi_-},& \ket{K \xi_-}
  \end{bmatrix}
  .
\end{equation}

\subsection{Analysis of adiabatic holonomies}
We examine the adiabatic holonomies of the quantum map.
Note that $\Delta$ and $\Theta$ depend only on $\mu$, $\lambda$, $\theta$ and
is the same ones for the quantum map
spin-$1/2$~(see, Eqs.~\eqref{eq:quantumMapSpin1/2},
and (\ref{eq:CosThetaQuantumMap1/2})).
Hence the degeneracy points in the parameter space and
the holonomy in the eigenvalues are the completely the same.
We will focus on the eigenspace holonomy in the following.

It is straightforward to obtain the gauge connection
from the eigenvectors. Since $\Theta$ depends on 
$\mu$, $\lambda$, $\theta$, the corresponding gauge connections are
defined through the derivative of $f$ by $\Theta$:
\begin{gather}
  \AF_{x}
  = i f^{\dagger}\pdfrac{}{x} f 
  = \frac{1}{2} \begin{bmatrix}0& -iI_2\\ +iI_2& 0\end{bmatrix}
  \pdfrac{\Theta}{x}
  ,
\end{gather}
for $x = \theta, \mu, \lambda$, where $I_2$ is the $2\times2$ unit matrix.
The corresponding 
Wilczek-Zee
%Mead-Truhlar-Berry 
gauge connections vanish to satisfy
the parallel transport condition, i.e..
$\AD_{\theta}=\AD_{\mu}=\AD_{\lambda}=0$.
For other gauge connections, $\mu$, $\lambda$, and $\theta$-dependences
are introduced through $\Theta$:
\begin{subequations}
\begin{align}
  \AF_{\eta} &
  =\frac{1}{2}
  \begin{bmatrix}
    -\sigma_y\cos\Theta
    &\sigma_y\sin\Theta
    \\
    \sigma_y\sin\Theta
    &
    \sigma_y\cos\Theta\end{bmatrix}
  ,\\
  \AF_{\varphi}&
  =
  \frac{1}{2}
  \begin{bmatrix}
    \sigma_z\cos\Theta&-\sigma_z\sin\Theta
    \\{}
    -\sigma_z\sin\Theta&-\sigma_z\cos\Theta
  \end{bmatrix}\cos\eta
  %\nonumber\\{}&\qquad
  +\frac{1}{2}\begin{bmatrix}\sigma_x&0\\{}0&\sigma_x\end{bmatrix}\sin\eta
  ,\\
  \AF_{\chi}&
  =\frac{1}{2}\begin{bmatrix}\sigma_z&0\\{}0&\sigma_z\end{bmatrix}\cos\eta
  %\nonumber\\{}&\qquad
  +\frac{1}{2}
  \begin{bmatrix}
    \sigma_x\cos\Theta&-\sigma_x\sin\Theta
    \\{}
    -\sigma_x\sin\Theta&-\sigma_x\cos\Theta
  \end{bmatrix}\sin\eta
  ,
\end{align}
\end{subequations}
% \begin{subequations}
% \begin{align}
%   \AF_{\eta} &
%   =\frac{1}{2}
%   \begin{bmatrix}
%     -\sigma_y\cos\Theta &\sigma_y\sin\Theta\\
%     \sigma_y\sin\Theta&\sigma_y\cos\Theta 
%   \end{bmatrix}
%   ,\\
%   \AF_{\varphi}&
%   =\frac{1}{2}
%   \begin{bmatrix}
%     \sigma_z\cos\Theta\cos\eta+\sigma_x\sin\eta& -\sigma_z\sin\Theta\cos\eta\\
%     -\sigma_z\sin\Theta\cos\eta&-\sigma_z\cos\Theta\cos\eta+\sigma_x\sin\eta
%   \end{bmatrix}
%   ,\\
%   \AF_{\chi}&
%   =\frac{1}{2}
%   \begin{bmatrix}
%     \sigma_z\cos\eta+\sigma_x\cos\Theta\sin\eta&
%     -\sigma_x\sin\Theta\sin\eta\\
%     -\sigma_x\sin\Theta\sin\eta&
%     \sigma_z\cos\eta-\sigma_x\cos\Theta\sin\eta
%   \end{bmatrix}
%   ,
% \end{align}
% \end{subequations}
and
\begin{subequations}
\begin{align}
  \AD_{\eta} &
  =\frac{1}{2}
  \begin{bmatrix}-\sigma_y&0\\{}0&\sigma_y\end{bmatrix}
  \cos\Theta
  ,\\
  \AD_{\varphi}&
  =
  \frac{1}{2}
  \begin{bmatrix}\sigma_z&0\\{}0&-\sigma_z\end{bmatrix}\cos\Theta\cos\eta
  %%\nonumber\\{}&\qquad
  +\frac{1}{2}\begin{bmatrix}\sigma_x&0\\{}0&\sigma_x\end{bmatrix}\sin\eta
  ,\\
  \AD_{\chi}&
  =\frac{1}{2}\begin{bmatrix}\sigma_z&0\\{}0&\sigma_z\end{bmatrix}\cos\eta
  %%\nonumber\\{}&\qquad
  +\frac{1}{2}\begin{bmatrix}\sigma_x&0\\{}0&-\sigma_x\end{bmatrix}
  \cos\Theta\sin\eta
  .
\end{align}
\end{subequations}
% \begin{subequations}
% \begin{align}
%   \AD_{\eta} &
%   =\frac{1}{2} 
%   \begin{bmatrix}-\sigma_y&0\\0&\sigma_y\end{bmatrix}\cos\Theta 
%   ,\\
%   \AD_{\varphi} &
%   =\frac{1}{2}
%   \begin{bmatrix}
%     \sigma_z\cos\Theta\cos\eta+\sigma_x\sin\eta& 0\\
%     0& -\sigma_z\cos\Theta\cos\eta+\sigma_x\sin\eta
%   \end{bmatrix}
%   ,\\
%   \AD_{\chi} &
%   =\frac{1}{2}
%   \begin{bmatrix}
%     \sigma_z\cos\eta+\sigma_x\cos\Theta\sin\eta& 0\\
%     0&\sigma_z\cos\eta-\sigma_x\cos\Theta\sin\eta
%   \end{bmatrix}
%   .
% \end{align}
% \end{subequations}

It is straightforward to evaluate the holonomy matrix 
$M(C_{\alpha})$~\eqref{eq:defM} for a closed loop $C_{\alpha}$
where $\alpha$(=$\mu,\lambda,\theta,\eta,\varphi,\chi$) is 
increased from $0$ to $2\pi$, once we take care the ``anholonomy''
in $\Theta$ along $C_{\alpha}$, which is also clarified in 
Section~\ref{sec:exampleAbelianCheon}.
For $\mu,\lambda,\theta$, we have
%\begin{subequations}
\begin{align}
  M(C_{\alpha})&
  = \exp\left(-\frac{i}{2}\begin{bmatrix}0& -iI_2\\ +iI_2& 0\end{bmatrix}
     \left.\Theta\right|_{\alpha=0}^{2\pi}\right)
  \nonumber\\ &
  = 
  \begin{bmatrix}
    I_2\cos\frac{\left.\Theta\right|_{\alpha=0}^{2\pi}}{2} 
    & -I_2\sin\frac{\left.\Theta\right|_{\alpha=0}^{2\pi}}{2} 
    \\ 
    +I_2\sin\frac{\left.\Theta\right|_{\alpha=0}^{2\pi}}{2} 
    & 
    I_2\cos\frac{\left.\Theta\right|_{\alpha=0}^{2\pi}}{2} 
  \end{bmatrix}
  ,
\end{align}
%\end{subequations}
where
\begin{subequations}
\begin{align}
  \left.\Theta\right|_{\theta=0}^{2\pi}&
  =
  \begin{cases}
    2\pi (-1)^{r/2}& \text{for $r$ is even}\\
    0& \text{for $r$ is odd}
  \end{cases}
  ,\\
  \left.\Theta\right|_{\lambda=0}^{2\pi}&
  = (-1)^{[\mu(2-q)/(2\pi)]} \pi (2-p)
  ,\\
  \left.\Theta\right|_{\mu=0}^{2\pi}&
  = (-1)^{[\lambda(2-p)/(2\pi)]} \pi (2-q)
  ,
\end{align}
\end{subequations}
and $r$ is defined in Eq.~\eqref{eq:defindexr}.
%% AT-2009-02
Hence $M(C_{\theta})$ exhibits only
Herzberg and Longuet-Higgins' sign change~\cite{Herzberg:DFS-35-77}
\begin{align}
  M(C_{\theta})& = (-1)^{1+r} I_4.
\end{align}
Also, the same kind of sign change appears along 
$C_{\mu}$ ($C_{\lambda}$) with even $p$ ($q$), 
i.e.,
$M(C_{\mu})=(-1)^{1+p/2} I_4$ ($M(C_{\lambda})=(-1)^{1+q/2} I_4$).
On the other hand, a mixture of the eigenspace holonomy 
and the Herzberg and Longuet-Higgins' sign change
occurs
along $C_{\mu}$ with odd $p$
\begin{align}
  M(C_{\mu})&
  =
  (-1)^{[\mu(2-q)/(2\pi)]+(p-1)/2}
  \begin{bmatrix}
    0 & -I_2\\ 
    I_2& 0
  \end{bmatrix}
  ,
\end{align}
and, along $C_{\lambda}$ with odd $q$
\begin{align}
  M(C_{\lambda})&
  =
  (-1)^{[\lambda(2-p)/(2\pi)]+(q-1)/2}
  \begin{bmatrix}
    0 & -I_2\\ 
    I_2& 0
  \end{bmatrix}
  ,
\end{align}
which do not incorporate mixing within
the degenerate eigenspaces.
%%
% \begin{subequations}
% \begin{align}
%   M(C_{\theta})& = (-1)^{1+r} I_4,
%   \\
%   M(C_{\mu})&
%   =
%   \begin{cases}
%     (-1)^{1+p/2} I_4& \text{for $p$ is even}\\
%     (-1)^{[\mu(2-q)/(2\pi)]+(p-1)/2}
%     \begin{bmatrix}
%       0 & -I_2\\ 
%       I_2& 0
%     \end{bmatrix}
%     & \text{for $p$ is odd}
%   \end{cases}
%   ,\\
%   M(C_{\lambda})&
%   =
%   \begin{cases}
%     (-1)^{1+q/2} I_4& \text{for $q$ is even}\\
%     (-1)^{[\lambda(2-p)/(2\pi)]+(q-1)/2}
%     \begin{bmatrix}
%       0 & -I_2\\ 
%       I_2& 0
%     \end{bmatrix}
%     & \text{for $q$ is odd}
%   \end{cases}
%   .
% \end{align}
% \end{subequations}
%%
% For  $C_{\mu}$ ($C_{\lambda}$) with odd $p$($q$), 
% we have a mixture of Herzberg and Longuet-Higgins' phase and
% the eigenspace holonomies, which do not incorporate mixing within
% the degenerate eigenspaces.
%%
Other holonomy matrices $M(C_{\alpha})$ ($\alpha=\eta,\varphi,\chi$) 
describes genuine Wilczek-Zee's phase holonomies:
%\begin{widetext}
\begin{subequations}
\begin{align}
  M(C_{\eta})&
  =
  \begin{bmatrix}
    \exp\left({i}\sigma_y\Omega_{\eta}/2\right)& 0\\
    0& \exp\left(-{i}\sigma_y\Omega_{\eta}/2\right)
  \end{bmatrix}
  ,\\
  M(C_{\varphi})&
  =
  \begin{bmatrix}
    \exp\left[-i(\sigma_z\cos\eta_1+\sigma_x\sin\eta_1)\Omega_{1}/2
      %%\pi\left(1-\beta_{\varphi}\right)
    \right]
    &0\\
    0&
    \exp\left[i(\sigma_z\cos\eta_1-\sigma_x\sin\eta_1)\Omega_{1}/2
      %%\pi\left(1-\beta_{\varphi}\right)
    \right]
  \end{bmatrix}
  ,\\
  M(C_{\chi})&
  =
  \begin{bmatrix}
    \exp\left[-i(\sigma_z\cos\eta_2+\sigma_x\sin\eta_2)\Omega_{2}/2
      %%\pi\left(1-\beta_{\chi}\right)
    \right]
    &0\\
    0&
    \exp\left[-i(\sigma_z\cos\eta_2-\sigma_x\sin\eta_2)\Omega_{2}/2
      %%\pi\left(1-\beta_{\chi}\right)
    \right]
  \end{bmatrix}
  ,
\end{align}
\end{subequations}
%\end{widetext}
% \begin{align}
%   M(C_{\eta})&
%   =
%   \begin{bmatrix}
%     e^{i\sigma_y\pi (1-\cos\Theta)}& 0\\
%     0& e^{-i\sigma_y\pi (1-\cos\Theta)}
%   \end{bmatrix}
%   ,\\
%   M(C_{\varphi})&
%   =
%   \begin{bmatrix}
%     \exp\left[-i(\sigma_z\cos\eta_1+\sigma_x\sin\eta_1)
%       \pi\left(1-\beta_{\varphi}\right)\right]
%     &0\\
%     0&
%     \exp\left[i(\sigma_z\cos\eta_1-\sigma_x\sin\eta_1)
%       \pi\left(1-\beta_{\varphi}\right)\right]
%   \end{bmatrix}
%   ,\\
%   M(C_{\chi})&
%   =
%   \begin{bmatrix}
%     \exp\left[-i(\sigma_z\cos\eta_2+\sigma_x\sin\eta_2)
%       \pi\left(1-\beta_{\chi}\right)\right]
%     &0\\
%     0&
%     \exp\left[-i(\sigma_z\cos\eta_2-\sigma_x\sin\eta_2)
%       \pi\left(1-\beta_{\chi}\right)\right]
%   \end{bmatrix}
%   .
% \end{align}
where
$\Omega_{\eta}\equiv 2\pi(1-\cos\Theta)$,
$\Omega_{j}\equiv2\pi (1 - \beta_{j})$,
$\eta_j
\equiv
-i\ln\left\{(\cos\Theta\cos\eta +i\sin\eta)/\beta_{j}\right\}$
for $j=1,2$,
% \begin{subequations}
% \begin{align}
%   \Omega_{\eta}
%   &\equiv 2\pi(1-\cos\Theta)
%   ,\\
%   \Omega_{1}
%   &\equiv2\pi (1 - \beta_{1})
%   ,&\quad
%   \eta_1&
%   \equiv
%   -i\ln\left\{(\cos\Theta\cos\eta +i\sin\eta)/\beta_{1}\right\}
%   ,\\
%   \Omega_{2}&
%   \equiv2\pi(1 - \beta_{2})
%   ,&\quad
%   \eta_2&
%   \equiv
%   -i\ln\left\{(\cos\Theta\cos\eta +i\sin\eta)/\beta_{2}\right\}
%   ,
% \end{align}
% \end{subequations}
% and
$\beta_{1}\equiv\sqrt{1 - (\sin\Theta\cos\eta)^2}$
and $\beta_{2}\equiv\sqrt{1 - (\sin\Theta\sin\eta)^2}$.

\section{Summary and outlook}
\label{sec:summary}

We have introduced a framework that is capable of describing
the eigenspace holonomy and the phase holonomy 
in a unified manner. Several examples have been shown.

In hindsight, it might seem rather odd that, in two decades
since the first discovery of Berry phase, the full, gauge
invariant formulation for quantum holonomy has not been
conceived prior to this work.
It should probably be attributed to the lack of the incentive
to improve on the original expression of Berry; 
Nothing other than the 
phase anholonomy has been anticipated 
for the adiabatic cyclic variation of
parameters for regular Hamiltonian system.
As a result,  the gauge invariant formulation 
must have seemed a redundant luxury.
However, if we once recognize the possibility of eigenstates' exchange
without level crossing for cyclic parameter variation, 
both for singular systems and for time-periodic systems, 
it becomes imperative
to treat the choice of basis frame explicitly within the formalism,
which has naturally lead us to arrive at the full, gauge invariant formulae.

In a simple minded view, it is the periodicity  of quasienergy, 
which is a result of the time-periodicity of the system,
that enables the eigenvalue holonomy in a natural manner.
For a Hamiltonian system, 
with the energy defined on entire real number, 
an energy eigenstate after cyclic variation of parameter 
cannot reach another eigenstate of different energy
in a usual way, since the crossing of levels is prohibited 
for adiabatic variation.
The only possible exception appears to be {\it singular} Hamiltonian systems, 
for which
the highest and the ground eigenenergy diverge~\cite{Cheon:PLA-248-285}.

In this work, we have focused on the most
elementary setting of quantum holonomy, {\it i.e.}, the adiabatic
excursion of pure quantum eigenstates 
along a closed path in the parameter space.
A vast amount of studies on the phase holonomy 
naturally suggests possible directions of extension of the present result. 
We mention only few of them.
A straightforward extension is to examine 
noncyclic path~\cite{Samuel:PRL-60-2339,Manini:PRL-85-3067}.
Also, loosening of the assumption of adiabaticity and resulting extension into
{\it e.g.} Aharonov-Anandan's nonadiabatic 
settings~\cite{Aharonov:PRL-58-1593}, are expected to be
straightforward thanks to 
the generality of 
Fujikawa's formulation~\cite{Fujikawa:PRD-72-025009}, 
which is the basis of our theory.
It seems timely as well as  interesting to examine the eigenspace holonomy 
in dissipative systems, for which we will need to horn appropriate techniques 
to treat of the eigenspace holonomy 
in mixed states~\cite{Uhlmann:RepMatPhys-24-229}.

Finally, we mention a question that is raised from the main 
result Eq.~\eqref{eq:defM}, which supplies a complete prescription to 
quantify the adiabatic quantum holonomy. How this helps to understand 
the exotic holonomies intuitively? 
Is it possible to find any 
underlying object or concept that governs them, for example, in the manner of 
diabolic point for the case of Berry phase? 
Herzberg and Longuet-Higgins has shown
that the phase holonomy along a closed loop $C$ implies 
the presence of spectral degeneracy in a surface $S$ enclosed by $C$
in the parameter space~\cite{Herzberg:DFS-35-77}
(see also, Refs.~\cite{Stone:PRSLA-351-141,Johansson:PRL-92-060406}).
Is there any counterpart of the argument of Herzberg and Longuet-Higgins 
for the exotic holonomies? 
This does not seem likely, for the case of exotic holonomy,
at the first glance, since there is no room to make $S$ from $C$ 
in all the examples shown in this paper.
We now believe that an affirmative answer is to be found
in the {\it complex parameter plane}, on which
we shall focus our attention in a forthcoming publication.

%%\begin{acknowledgments}
\section*{Acknowledgments}
This work has been partially supported by
the Grant-in-Aid for Scientific Research of  Ministry of Education,
Culture, Sports, Science and Technology, Japan
under the Grant number 18540384.
%%\end{acknowledgments}

\appendix

\section{The gauge theory for Hamiltonian time evolution}
\label{app:MforHamiltonian}

We will derive the holonomy matrix (Eq.~\eqref{eq:defM}) for 
a system described by Hamiltonian $\hat{H}(s)$
that depends on a time-dependent parameter $s$.
Suppose that $s$ is moved from $s'$ to $s''$ along a path $C$,
during $0 \le t \le T$.
The corresponding time evolution operator is
\begin{align}
  \hat{U}(\set{s_t}_{t\in[0,T]})
  \equiv\Texp\left(-\frac{i}{\hbar} \int_0^T \hat{H}(s_t)dt\right)
  .
\end{align}
We divide the time interval $[0,T]$ into $L$ parts.
Let $t_{l}\equiv (l/L)T$.
A short time evolution operator $\hat{U}_l$ is accordingly introduced:
\begin{align}
  \hat{U}_l
  \equiv\Texp\left(-\frac{i}{\hbar} \int_{t_l}^{t_{l+1}}
    \hat{H}(s_t)dt\right)
  .
\end{align}
When $L$ is large enough, we have
\begin{align}
  \hat{U}_l&
%   = 
%   1 
%   -\frac{i}{\hbar} \int_{t_l}^{t_{l+1}} \hat{H}(s_t)dt
%   + \mathcal{O}\left((T/L)^2\right)
%   = 
%   1 
%   -\frac{i}{\hbar} \hat{H}({s}_l)\frac{T}{L}
%   + \mathcal{O}\left((T/L)^2\right)
%   \\ &
  = 
  \exp\left(-\frac{i}{\hbar} \hat{H}({s}_l)\epsilon\right)
  + \mathcal{O}\left(\epsilon^2\right)
  %\exp\left\{-\frac{i}{\hbar} \hat{H}({s}_l)\frac{T}{L}\right\}
  %+ \mathcal{O}\left((T/L)^2\right)
  ,
\end{align}
where ${s}_l\equiv \frac{1}{2}(s_{t_{l+1}}+s_{t_l})$
%%AT2009-02
and $\epsilon\equiv T/L$.
Now we introduce a $s$-dependent unitary operator
\begin{align}
  \hat{U}_s \equiv 
  \exp\left(-\frac{i}{\hbar} \hat{H}({s})\epsilon\right)
  %\exp\left\{-\frac{i}{\hbar} \hat{H}({s})\frac{T}{L}\right\}
  .
\end{align}
Accordingly we have
\begin{align}
  \hat{U}(\set{s_t}_{t\in[0,T]})
  = \Tprod_{l=0}^{L-1}\;\hat{U}_{s_l} 
  + \mathcal{O}\left(\epsilon\right)
  %+ \mathcal{O}\left(T^2/L\right)
  .
\end{align}
Since we choose $L$ so as to satisfy $\epsilon \ll 1$, 
our formulation explained 
in Section~\ref{sec:unifiedTheory} is straightforwardly applicable.

First, for a given $f(s)$, the $f(s)$-representation of $U(s)$ is
\begin{align}
  Z(s)&
%  = \left\{f(s)^{\dagger}\right\}\times\hat{U}(s) \times f(s)
%   = 1 
%   -\frac{i}{\hbar}\left\{f(s)^{\dagger}\right\}\hat{H}({s})f(s)\frac{T}{L}
%   + \mathcal{O}\left((T/L)^2\right)
%  \\ &
  = \exp\left(
    -\frac{i}{\hbar}\left\{f(s)^{\dagger}\right\}\hat{H}({s})f(s)\epsilon
  \right)
  + \mathcal{O}\left(\epsilon^2\right)
%   = \exp\left(
%     -\frac{i}{\hbar}\left\{f(s)^{\dagger}\right\}\hat{H}({s})f(s)\frac{T}{L}
%   \right)
%   + \mathcal{O}\left((T/L)^2\right)
%%
%   \\ &
%   = \left\{f(s)^{\dagger}\right\}\exp\left\{ 
%     -\frac{i}{\hbar}\hat{H}({s})\frac{T}{L}
%   \right\}f(s)
%   + \mathcal{O}\left((T/L)^2\right)
  .
\end{align}
On the other hand, we have
\begin{align}
  &
  \Texp\left(i\int_{s_l}^{s_{l+1}} A(s)ds\right)
%  = 1 + i\int_{s_l}^{s_{l+1}} A(s)ds + \mathcal{O}((s_{l+1} - s_l)^2)
%  \\ &
%  = 1 + i A(\bar{s}_l)(s_{l+1} - s_l)
%  + \mathcal{O}((s_{l+1} - s_l)^2)
%\end{align}
%%$\bar{s}_l\equiv (s_{l+1} + s_l)/2$.
% \begin{align}
%   s_{l+1} - s_l &
%   = (s_{t_{l+2}}+s_{t_l+1})/2 - (s_{t_{l+1}}+s_{t_l})/2
%   = (s_{t_{l+2}} - s_{t_l})/2
%   \\ &
%   = \dot{s}\frac{T}{L} 
%   + \mathcal{O}\left(\left(\frac{T}{L}\right)^2\right)
% \end{align}
%%
%\begin{align}
%   \Texp\left\{i\int_{s_l}^{s_{l+1}} A(s)ds\right\}&
%   = 1 + i A(\bar{s}_l)\dot{s}\frac{T}{L} 
%   + \mathcal{O}\left(\left(\frac{T}{L}\right)^2\right)
%   = 1 + i A({s}_l)\dot{s}\frac{T}{L} 
%   + \mathcal{O}\left(\left(\frac{T}{L}\right)^2\right)
%   \\ &
  %\nonumber\\{}&
  = \exp\left(i A({s}_l)\dot{s}\epsilon\right)
  + \mathcal{O}\left(\epsilon^2\right)
%  = \exp\left(i A({s}_l)\dot{s}\frac{T}{L}\right)
%  + \mathcal{O}\left(\left(\frac{T}{L}\right)^2\right)
%
%   \\ &
%   = 1 
%   - \left\{f(s_l)^{\dagger}\right\}\pdfrac{f(s_l)}{s}\dot{s}\frac{T}{L} 
%   + \mathcal{O}\left(\left(\frac{T}{L}\right)^2\right)
%   \\ &
%   = \left.
%   \left\{f(s)^{\dagger}\right\}
%   \exp\left\{i \times i\dot{s}\pdfrac{}{s}\frac{T}{L}\right\}
%   f(s)\right|_{s = s_l}
%   + \mathcal{O}\left(\left(\frac{T}{L}\right)^2\right)
  ,
\end{align}
where we assumed $\dot{s} = \mathcal{O}(1)$.
Hence we have
\begin{align}
  \Texp\left(i\int_{s_l}^{s_{l+1}} A(s)ds\right)Z(s_l)
  &
%   \\ &
%   = \exp\left\{i A({s}_l)\dot{s}\frac{T}{L}\right\}
%   \times\exp\left\{-\frac{i}{\hbar}\left\{f(s_l)^{\dagger}\right\}
%     \hat{H}({s_l})f(s_l)\frac{T}{L}\right\}
%   + \mathcal{O}\left((T/L)^2\right)
%   \\ &
%   =
%   \exp\left\{-\frac{i}{\hbar}
%     \left(\left\{f(s_l)^{\dagger}\right\}\hat{H}({s_l})f(s_l)
%       - \hbar A({s}_l)\dot{s}\right)\frac{T}{L}\right\}
%   + \mathcal{O}\left((T/L)^2\right)
   =
   \exp\left(-\frac{i}{\hbar}F(s_l, \dot{s})\epsilon\right)
   %\nonumber\\{}&\qquad\qquad{}
   + \mathcal{O}\left(\epsilon^2\right)
%   \exp\left(-\frac{i}{\hbar}F(s_l, \dot{s})\frac{T}{L}\right)
%   + \mathcal{O}\left((T/L)^2\right)
   ,
\end{align}
where we introduce Fujikawa's Hamiltonian
matrix~\cite{Fujikawa:PRD-72-025009} 
\begin{align}
  F(s, \dot{s})
  \equiv \left\{f(s)^{\dagger}\right\}\hat{H}({s})f(s) - \hbar A({s})\dot{s}
 .   
\end{align}
In the limit $L\to\infty$, the effective time evolution operator 
for Fujikawa's formulation~\cite{Fujikawa:PRD-72-025009} is
% \begin{align}
%   B_{\rm d}(\set{s_l}_{l=0}^{L})&
%   = \Tprod_{l=0}^{L-1}
%   \exp\left\{-\frac{i}{\hbar} F(s_l, \dot{s})\right\}
%   + \mathcal{O}\left(T^2/L\right)
% \end{align}
\begin{align}
  B_{\rm d}(\set{s_t})&
  = \Texp\left(-\frac{i}{\hbar}\int_0^{T}F(s_t, \dot{s}_t) dt\right)
  .
\end{align}

Next, let us examine the adiabatic change of $s$.
Let $f(s)$ be an adiabatic basis for $\hat{H}(s)$.
Now 
\begin{align}
  H^{\rm D}(s)\equiv
  f(s)^{\dagger}\hat{H}({s})f(s) 
\end{align}
is a diagonal matrix whose non-zero elements are 
the eigenvalues of $\hat{H}(s)$.
Thanks to the adiabatic theorem, 
we employ the diagonal approximation for $B_{\rm d}(\set{s_t})$:
\begin{align}
  B_{\rm d}(\set{s_t})&
  \simeq
  B_{\rm ad}(\set{s_t})
  \equiv
  \Texp\left(-\frac{i}{\hbar}\int_0^{T}
    F^{\rm D}(s,\dot{s})dt\right)
  ,
\end{align}
where $F^{\rm D}(s,\dot{s})$ is defined as
\begin{align}
  F^{\rm D}(s,\dot{s})\equiv H^{\rm D}(s) - \hbar \AD({s})\dot{s}
  .
\end{align}
Hence $B_{\rm ad}(\set{s_t})$ is decomposed into
geometric and dynamical factors:
\begin{align}
  B_{\rm ad}(\set{s_t})
  &
  = 
  \Texp\left(i \int_{s'}^{s''} A^{\rm D}({s})ds\right)
  %\nonumber\\{}&\qquad\times
  \exp\left(-\frac{i}{\hbar}\int_0^{T} H^{\rm D}(s)dt\right)
  .
\end{align}
Now it is trivial to apply our formulation explained in the main text
to obtain the holonomy matrix~(Eq.~(\ref{eq:defM})).

\section{A derivation of Eq.~(\ref{eq:MofCMinQuamtumMap1/2})}
\label{app:meridianKickedTop1/2}

To evaluate Eq.~\eqref{eq:WofCMinQuamtumMap1/2}, 
we need to obtain
the image of $C_{\theta}$ in the sphere $(\Theta, \varphi)$.
First, we remind that $\Theta$ is a periodic function of
$\theta$ from Eqs.~\eqref{eq:DeltaQuantumMap1/2} and
\eqref{eq:CosThetaQuantumMap1/2}.
Hence $\left.\Theta\right|_{\theta=0}^{2\pi}$ must be a multiple of $2\pi$.
Second, we remind that there is no spectrum degeneracy along the path
$C_{\theta}$ if $r_{\pm}\notin\Integer$, where
$r_{\pm}\equiv \left\{\lambda(2-p)\pm\mu(2-q)\right\}/(2\pi)$.
Namely, the nondegenerate regions are divided into
squares by the lattice $(r_+, r_-)\in\Integer^2$.
%With a pair of integers $(a,b)$, we specify a square
%$\set{(p_+, p_-)\vert a < p_+ < a+1, b < p_- < b+1}$
%in the nondegeneracy regions.
Within each square, 
$\left.\Theta\right|_{\theta=0}^{2\pi}$ takes a constant value.
Accordingly, it is suffice to evaluate them
at representative points of the squares.

Let us examine the case $(r_+ - r_-)/2$ is an integer $k$,
which implies
$\Delta = 2\cos^{-1}((-1)^k \cos(\pi(r_+ + r_-)/2)$,
$\cos\Theta 
= (-1)^k\cos\theta\sin(\pi(r_+ + r_-)/2)/\sin(\Delta/2)$,
and
$\sin\Theta 
= \sin\theta\sin(\pi(r_+ + r_-)/2)/\sin(\Delta/2)$.
Accordingly we have
\begin{align}
  e^{i\Theta} 
  = (-1)^{[k+(r_+ + r_-)/2]} \exp\{i (-1)^k\theta\}
  ,
\end{align}
which implies the image of $C_{\theta}$ is also a great meridian loop 
in the sphere $(\Theta, \varphi)$.
Thus we have
\begin{align}
  \left.\Theta\right|_{\theta=0}^{2\pi} 
  = (-1)^k 2\pi
  .
\end{align}
This result is also valid for the case that
$([r_+] - [r_-])/2$ is an integer $k$.

Let us examine the case $(r_+ + r_-)/2$ is an integer $l$,
which implies
$\Delta = 2\cos^{-1}((-1)^l \cos(\pi(r_+ - r_-)/2)$,
$\cos\Theta 
= (-1)^{l+[(r_+ - r_-)/2]}$
and
$\sin\Theta 
= 0$.
Namely, the image of $C_{\theta}$ is a point in $(\Theta,\varphi)$ space.
Accordingly we have
\begin{align}
  \left.\Theta\right|_{\theta=0}^{2\pi} 
  = 0
  .
\end{align}
This result is also valid for the case that
$[r_+] - [r_-]$ is an odd integer.

To summarize the argument above,
we have, from Eq.~\eqref{eq:WofCMinQuamtumMap1/2}, 
\begin{align}
  M(C_{\theta})
  =
  \exp\left\{i\pi
    \left(1 + [r_+] - [r_-]\right)\right\}
  ,
\end{align}
which is equivalent with Eq.~\eqref{eq:MofCMinQuamtumMap1/2}.

\section{Algebraic properties of  $\hat\tau_{\alpha}$}
\label{app:AlgebraicPropsTau}
We summarize the algebraic properties of $\hat\tau_{\alpha}$.
Details are found in, for example, Ref.~\cite{Avron:CMP-124-595}.
It is straightforward to show that they form 
a Clifford algebra~\eqref{eq:CliffordAlgebra}, which implies
the following formulas:
\begin{align}
  \hat\tau_{\alpha}\hat\tau_{\beta}\hat\tau_{\alpha}
  = 2\delta_{\alpha\beta}\hat\tau_{\alpha} - \hat\tau_{\beta}
  .
\end{align}
For $B_{\alpha}\in\Real$, 
\begin{align}
  \left(\sum_{\alpha}B_{\alpha}\hat\tau_{\alpha}\right)^2 
  = \left\|B\right\|^2,
\end{align}
where $\left\|B\right\|$ is a norm of real vector, i.e.
$\left\|B\right\|
= \sqrt{\sum_{\alpha}B_{\alpha}^2}$.
Furthermore, for a real unit vector $n$ that satisfies $\|n\|=1$,
\begin{align}
  \exp\left(-i\lambda\sum_{\alpha}n_{\alpha}\hat\tau_{\alpha}\right)
  = \cos\lambda - i 
  \left(\sum_{\alpha}n_{\alpha}\hat\tau_{\alpha}\right)\sin\lambda
  ,
\end{align}
which is $2\pi$-periodic in $\lambda$.
Next formula is useful to investigate quantum maps:
\begin{align}
  \label{eq:tauSymmetricProduct}
  &
  \exp\left(-i\lambda\hat\tau_{\alpha}\right)
  \hat\tau_{\beta}
  \exp\left(-i\lambda\hat\tau_{\alpha}\right)
  \nonumber\\{}&
  = (1 - \delta_{\alpha\beta})\hat\tau_{\beta}
  + \delta_{\alpha\beta} \{\cos(2\lambda)\hat\tau_{\beta}
  - i\sin(2\lambda)\}
  .
\end{align}
Finally, we show a representation of $\hat\tau_{\alpha}$ by
complex $4\times4$ matrices $\tau_{\alpha}$
in terms of complex $2\times2$ matrices defined 
in Eq.~\eqref{eq:def2time2matrices}, i.e.,
$\hat\tau_{\alpha} = f_0 \tau_{\alpha} f_0^{\dagger}$,
where %%$f_0$ is a frame that is a sequence of basis vector
$
  f_0\equiv 
  \begin{bmatrix}
    \ket{e_1},& \ket{\TR e_1},&\ket{e_2},& \ket{\TR e_2}
  \end{bmatrix}
$:
\begin{equation}
\begin{gathered}
  \tau_0=\begin{bmatrix}I_2&0\\{}0&-I_2
%     1&0&&\\
%     0&1&&\\
%     &&-1&0\\
%     &&0&-1
  \end{bmatrix}
  ,
  \quad%\\
  \tau_1=\begin{bmatrix}0& i\sigma_y\\ -i\sigma_y& 0
%     &&0&1\\
%     &&-1&0\\
%     0&-1&&\\
%     1&0&&
  \end{bmatrix}
  ,\quad
  \tau_2=\begin{bmatrix}0& -i\sigma_x\\ i\sigma_x&0
%     &&0&-i\\
%     &&-i&0\\
%     0&i&&\\
%     i&0&&
  \end{bmatrix}
  ,\\ 
  %%,\quad
  \tau_3=\begin{bmatrix}0&I_2\\ I_2&0
%     &&1&0\\
%     &&0&1\\
%     1&0&&\\
%     0&1&&
  \end{bmatrix}
  ,\quad
  \tau_4=\begin{bmatrix}0& -i\sigma_z\\ i\sigma_z&0
%     &&-i&0\\
%     &&0&i\\
%     i&0&&\\
%     0&-i&&
  \end{bmatrix}
  .
\end{gathered}
% \begin{gathered}
%   \tau_0=\begin{bmatrix}
%     1&0&&\\
%     0&1&&\\
%     &&-1&0\\
%     &&0&-1
%   \end{bmatrix}
%   ,\quad
%   \tau_1=\begin{bmatrix}
%     &&0&1\\
%     &&-1&0\\
%     0&-1&&\\
%     1&0&&
%   \end{bmatrix}
%   ,\quad
%   \tau_2=\begin{bmatrix}
%     &&0&-i\\
%     &&-i&0\\
%     0&i&&\\
%     i&0&&
%   \end{bmatrix}
%   ,\\ %%,\quad
%   \tau_3=\begin{bmatrix}
%     &&1&0\\
%     &&0&1\\
%     1&0&&\\
%     0&1&&
%   \end{bmatrix}
%   ,\quad
%   \tau_4=\begin{bmatrix}
%     &&-i&0\\
%     &&0&i\\
%     i&0&&\\
%     0&-i&&
%   \end{bmatrix}
%   .
% \end{gathered}
\end{equation}

\section{A derivation of Eq.~(\ref{eq:QuantumMap3/2Tilde})}
\label{app:QM3/2Tilde}
We derive Eq.~(\ref{eq:QuantumMap3/2Tilde})
from Eq.~\eqref{eq:quantumMapSpin3/2}.
First, let $\Bmu\equiv\mu(2-q)/2$ and $\Blambda\equiv\lambda(2-p)/2$.
Hence we have
\begin{align}
  \hat{U}
  &
%   \equiv
%   \exp\left\{-i\frac{\mu}{2}\left(\frac{q}{2} 
%       + \frac{2-q}{2}\hat\tau_0\right)\right\}
%   \exp\left\{-i\lambda\left(\frac{p}{2} 
%       + \frac{2-p}{2}\sum_{\alpha=0}^4
%       n_{\alpha}\hat\tau_{\alpha}\right)\right\}
%   \nonumber\\ &\qquad
%   {}\times
%   \exp\left\{-i\frac{\mu}{2}\left(\frac{q}{2} 
%       + \frac{2-q}{2}\hat\tau_0\right)\right\}
%  ,\\ &
  =
  e^{-i(\mu q + \lambda p)/2}
  e^{-i\Bmu\hat\tau_0/2}
  e^{-i\Blambda\sum_{\alpha} n_{\alpha}\hat\tau_{\alpha}}
  e^{-i\Bmu\hat\tau_0/2}
  ,
  \nonumber\\ &
%   =
%   e^{-i(\mu q + \lambda p)/2}
%   e^{-i\Bmu\hat\tau_0/2}
%   \left(
%     \cos\Blambda -i\sum_{\alpha}n_{\alpha}\hat\tau_{\alpha}\sin\Blambda
%   \right)
%   e^{-i\Bmu\hat\tau_0/2}
%  ,\\ &
  =
  e^{-i(\mu q + \lambda p)/2}e^{-i\Bmu\hat\tau_0}
  \Bigl(
    \cos\Blambda
    %\Bigr.\nonumber\\{}&\qquad\qquad\Bigl.
    -i\sum_{\alpha}n_{\alpha}
    e^{-i\Bmu\hat\tau_0/2}
    \hat{\tau}_{\alpha}e^{-i\Bmu\hat\tau_0/2}
    \sin\Blambda
    \Bigr)
  %% From \ref{eq:tauSymmetricProduct}
  ,\nonumber\\ &
%   =
%   e^{-i(\mu q + \lambda p)/2}
%   \left(\left(\cos\Bmu -i\tau_0\sin\Bmu\right)\cos\Blambda
%     -in_{0}
%     \left(\tau_0\cos\Bmu - i\sin\Bmu\right)
%     \sin\Blambda
%     -i\sum_{\alpha\ne0}n_{\alpha}\hat{\tau_{\alpha}}
%     \sin\Blambda
%   \right)
%   ,\\ &
  =
  e^{-i(\mu q + \lambda p)/2}
  \left(
    k -i\sum_{\alpha\ne0}\tilde{l}_{\alpha}\hat{\tau}_{\alpha}
  \right)
  ,
\end{align}
where we used Eq.~\eqref{eq:tauSymmetricProduct} above, and,
\begin{equation}
\begin{aligned}
  k&
  \equiv\cos\Bmu\cos\Blambda - n_0\sin\Bmu\sin\Blambda
  ,\\ 
  \tilde{l}_0&
  \equiv\sin\Bmu\cos\Blambda+n_0\cos\Bmu\sin\Blambda
  ,\\ 
  \tilde{l}_{\alpha(\ne0)}&
  \equiv n_{\alpha}\sin\Blambda
  .
\end{aligned}
\end{equation}
Hence we arrive Eq.~(\ref{eq:QuantumMap3/2Tilde}).

\section{Diagonalization of 
  $\sum_{\alpha} n_{\alpha}\hat{\tau}_{\alpha}$}
\label{app:DiagonalizationOfN}

Let $n_{\alpha}$ ($\alpha=0,\dots,4$) be real, and, normalized, i.e.
$\sum_{\alpha=0}^4 n_{\alpha}^2 = 1$.
We will obtain the eigenvectors of
\begin{align}
  \hat\tau(\set{n_{\alpha}})
  \equiv\sum_{\alpha=0}^4 n_{\alpha}\hat{\tau}_{\alpha}
  ,
\end{align}
with the help of the quaternionic structure of the Hilbert space 
induced by the fermion time reversal invariance~\cite{Avron:CMP-124-595}.
We follow the convention of quaternions explained in 
Ref~\cite{Avron:CMP-124-595},
e.g.,
$\qi^2 = \qj^2 = \qk^2 = -1$, 
$\qi\qj = -\qj\qi = \qk$,
$\qj\qk = -\qk\qj = \qi$,
and $\qk\qi = -\qi\qk = \qj$.
% \begin{equation}
% \begin{gathered}
%   \qi^2 = \qj^2 = \qk^2 = -1\\
%   \qi\qj = -\qj\qi = \qk
%   ,\quad
%   \qj\qk = -\qk\qj = \qi
%   ,\quad
%   \qk\qi = -\qi\qk = \qj
%   .
% \end{gathered}
% \end{equation}

In the Hilbert space $\Hilbert$ of spin-$\frac{3}{2}$, 
we employ
a right quaternionic action for $\ket\psi\in\Hilbert$:
%\begin{align}
$\ket{\psi}\qi \equiv i\ket\psi$%
, and,
$
\ket{\psi}\qj \equiv \hat\TR\left(\ket\psi\right)%
$%
,
%\end{align}
which implies 
%\begin{align}
$
  \ket{\psi}\qk = \ket{\psi}(\qi\qj) = \left(\ket{\psi}\qi\right)\qj
  = \hat\TR\left(\ket{\psi}\qi\right)
  = -i\left\{ \hat\TR\left(\ket{\psi}\right)\right\}%
$
.
%\end{align}
Accordingly, for  $z_j, w_j \in\Complex$, we have
%\begin{subequations}
\begin{align}
  \ket\psi &
  = z_1 \ket{e_1} + w_1 \ket{\TR e_1}
  +  z_2 \ket{e_2} + w_2 \ket{\TR e_2} 
  %= \ket{e_1} z_1 + (\ket{e_1}\qj) w_1
  %+ \ket{e_2} z_2 + (\ket{e_2}\qj) w_2
  \nonumber\\ &
  = \ket{e_1} (z_1 + \qj w_1) + \ket{e_2} (z_2 + \qj w_2)
  \nonumber\\ &
  = f_{\rm q}
  \begin{bmatrix}
    z_1 + \qj w_1\\ z_2 + \qj w_2
  \end{bmatrix}
  ,
\end{align}
%\end{subequations}
where
\begin{align}
  f_{\rm q}\equiv \begin{bmatrix}\ket{e_1}, &\ket{e_2}\end{bmatrix}
\end{align}
is the standard frame of the quaternionic Hilbert space.
Hence we obtain a natural correspondence 
between four-dimensional complex vector space 
and two-dimensional quaternionic vector space.
This induces a representation of $\hat\tau_{\alpha}$ 
by $2\times2$ quaternionic matrix $\tau^{\rm q}_{\alpha}$, i.e.,
\begin{align}
  \hat\tau_{\alpha}
  = f_{\rm q}\; \tau^{\rm q}_{\alpha}\; f_{\rm q}^{\dagger}
  ,
\end{align}
where
\begin{equation}
  \begin{gathered}
  \tau^{\rm q}_0
  = \begin{bmatrix}1&0\\0&-1\end{bmatrix}
  ,\quad
  \tau^{\rm q}_1\equiv\begin{bmatrix}0&-\qj\\ \qj&0\end{bmatrix}
  ,\quad
  \tau^{\rm q}_2\equiv\begin{bmatrix}0&\qk\\ -\qk&0\end{bmatrix}
  ,
  \\
  \tau^{\rm q}_3\equiv\begin{bmatrix}0&1\\1&0\end{bmatrix}
  ,\quad
  \tau^{\rm q}_4\equiv\begin{bmatrix}0&-\qi\\ \qi&0\end{bmatrix}
  .
  \end{gathered}
\end{equation}

We introduce spherical variables $\theta$, $\eta$, $\chi$ and $\varphi$
to parameterize $n_{\alpha}$ as
$n_0=\cos\theta$,
$n_1=\cos\chi\sin\eta\sin\theta$, 
$n_2=\sin\chi\sin\eta\sin\theta$
$n_3=\cos\varphi\cos\eta\sin\theta$,
and
$n_4=\sin\varphi\cos\eta\sin\theta$.
% \begin{equation}
%   \begin{gathered}
%   n_0=\cos\theta
%   ,\quad
%   \begin{bmatrix} n_1\\ n_2\end{bmatrix}
%   = \begin{bmatrix}\cos\chi\\ \sin\chi\end{bmatrix}\sin\eta\sin\theta
%   ,
%   \\
%   \begin{bmatrix} n_3\\ n_4\end{bmatrix}
%   = \begin{bmatrix}\cos\varphi\\ \sin\varphi\end{bmatrix}\cos\eta\sin\theta
%   .
%   \end{gathered}
% \end{equation}
In terms of $2\times2$ quaternionic matrix, we have
%\begin{widetext}
\begin{align}
  \hat\tau(\set{n_{\alpha}})
  &
  = 
  f_{\rm q}
  \begin{bmatrix}
    \cos\theta& 
    (\cos\eta - \qj e^{\qi(\varphi+\chi)}\sin\eta)e^{-\qi\varphi}\sin\theta\\
    e^{+\qi\varphi}(\cos\eta + \qj e^{\qi(\varphi+\chi)}\sin\eta)\sin\theta&
    -\cos\theta
  \end{bmatrix}
  f_{\rm q}^{\dagger}
  ,
\end{align}
%\end{widetext}
where we introduce
\begin{align}
  h\equiv\qj e^{\qi(\varphi+\chi)}
  .
\end{align}
Since 
$
{h}^2 
= \qj e^{\qi(\varphi+\chi)}e^{-\qi(\varphi+\chi)}\qj 
= -1
$, 
we have $e^{h\eta} = \cos\eta + h\sin\eta$.
Hence we have
\begin{align}
  \hat\tau(\set{n_{\alpha}})
  &
  = 
  f_{\rm q}
  \begin{bmatrix}
    \cos\theta& e^{-h\eta}e^{-\qi\varphi}\sin\theta \\
    e^{+\qi\varphi} e^{h\eta}\sin\theta& -\cos\theta
  \end{bmatrix}
  f_{\rm q}^{\dagger}
  .
\end{align}
Now it is straightforward to see the eigenvalues of
$\hat\tau(\set{n_{\alpha}})$ are $\pm 1$.
Let $\ket{\xi_{\pm}^{(0)}}$ be corresponding eigenvectors:
\begin{subequations}
\begin{align}
  \ket{\xi_+^{(0)}}&
  \equiv 
  f_{\rm q}
  \begin{bmatrix}
    e^{-h\eta/2}\cos\frac{\theta}{2}\\
    e^{i\varphi}e^{h\eta/2}\sin\frac{\theta}{2}
  \end{bmatrix}
  ,
  \\
  \ket{\xi_-^{(0)}}&
  \equiv 
  f_{\rm q}
  \begin{bmatrix}
    e^{-h\eta/2}\left(-\sin\frac{\theta}{2}\right)\\
    e^{i\varphi}e^{h\eta/2}\cos\frac{\theta}{2}
  \end{bmatrix}
  .
\end{align}
\end{subequations}
Instead of the two above, we put a phase factor on them in the following:
\begin{align}
  \ket{\xi_{\pm}}&
  \equiv\ket{\xi_{\pm}^{(0)}} e^{-i(\varphi+\chi)/2}
  ,
\end{align}
where we need to take care about the noncommutativity of multiplication 
in quaternions.
In the complex Hilbert space, $\ket{\xi_{\pm}}$ are expressed as
\begin{subequations}
\begin{align}
  \ket{\xi_+}&
  = \ket{d_1}\cos\frac{\theta}{2} + \ket{d_2}\sin\frac{\theta}{2}
  ,
  \\
  \ket{\xi_-}&
  = \ket{d_1}\left(-\sin\frac{\theta}{2}\right) 
  + \ket{d_2}\cos\frac{\theta}{2}
  ,
\end{align}
\end{subequations}
where $\ket{d_{1}}$ and $\ket{d_{2}}$ are orthonormal
\begin{subequations}
\begin{align}
  \ket{d_1}&
  \equiv \ket{e_1}\left(e^{-i(\varphi+\chi)/2}\cos\frac{\eta}{2}\right)
  %\nonumber\\{}&\qquad
  +\ket{\TR e_1}\left(-e^{+i(\varphi+\chi)/2}\sin\frac{\eta}{2}\right)
  ,
  \\
  \ket{d_2}&
  \equiv
  \ket{e_2}\left(e^{i(\varphi-\chi)/2}\cos\frac{\eta}{2}\right)
  %\nonumber\\{}&\qquad
  +\ket{\TR e_2}\left(e^{-i(\varphi-\chi)/2}\sin\frac{\eta}{2}\right)
  .
\end{align}
\end{subequations}
We explain the rest of 
eigenvectors of 
$\hat\tau(\set{n_{\alpha}})$ for the complex Hilbert space.
They are obtained by the time-reversal operation on 
$\ket{\xi_{\pm}}$:
\begin{subequations}
\begin{align}
  \ket{\TR \xi_+}&
  = \ket{\TR d_1}\cos\frac{\theta}{2} + \ket{\TR d_2}\sin\frac{\theta}{2}
  ,
  \\
  \ket{\TR \xi_-}&
  = \ket{\TR d_1}\left(-\cos\frac{\theta}{2}\right) 
  + \ket{\TR d_2}\cos\frac{\theta}{2}
  ,
\end{align}
\end{subequations}
where
\begin{subequations}
\begin{align}
  \ket{\TR d_1}&
  = \ket{e_1}\left(+e^{-i(\varphi+\chi)/2}\sin\frac{\eta}{2}\right)
  %\nonumber\\{}&\qquad
  + \ket{\TR e_1}\left(e^{+i(\varphi+\chi)/2}\cos\frac{\eta}{2}\right)
  ,
  \\
  \ket{\TR d_2}&
  = 
  \ket{e_2}\left(-e^{+i(\varphi-\chi)/2}\sin\frac{\eta}{2}\right)
  %\nonumber\\{}&\qquad
  +\ket{\TR e_2}\left(e^{-i(\varphi-\chi)/2}\cos\frac{\eta}{2}\right)
  .
\end{align}
\end{subequations}
Because of $\hat{K}^2 = -1$,
$\set{\ket{\xi_{\pm}},\ket{K \xi_{\pm}}}$ 
is a complete orthogonal system for $\Hilbert$.

To compute gauge connections for 
$\set{\ket{\xi_{\pm}},\ket{K \xi_{\pm}}}$, 
it is useful to summarize the basis transformation between
\begin{subequations}
\begin{align}
  f\equiv &
  \begin{bmatrix}
    \ket{\xi_+},& \ket{K \xi_+},&\ket{\xi_-},& \ket{K \xi_-}
  \end{bmatrix}
  ,
  \intertext{and}
  f_0\equiv& 
  \begin{bmatrix}
    \ket{e_1},& \ket{K e_1},&\ket{e_2},& \ket{K e_2}
  \end{bmatrix}
  .
\end{align}
\end{subequations}
It is straightforward to obtain the following
\begin{align}
  f 
  &
  = f_0 
  \exp\left(- \frac{i}{2} g_{4} \chi\right) 
  \exp\left(- \frac{i}{2} g_{3} \varphi\right) 
  %\nonumber\\{}&\qquad\times
  \exp\left(- \frac{i}{2} g_{2} \eta\right) 
  \exp\left(- \frac{i}{2} g_{1} \theta\right) 
  %e^{-i g_{3} \varphi/2}
  %e^{-i g_{2} \eta/2}e^{-i g_{1} \theta/2}
  %G_4(\chi)G_3(\varphi)G_2(\eta) G_1(\theta)
\end{align}
where 
% $G_{\alpha}(x)$ are unitary matrices
% \begin{align}
%   G_{\alpha}(x) = e^{-i g_{\alpha} x/2}
%   ,
% \end{align}
% and 
$g_{\alpha}$ are $4\times4$ Hermite matrices
\begin{equation}
\begin{gathered}
  g_1 
  % &
  \equiv 
  \begin{bmatrix}
    0& -iI_2\\ iI_2& 0
  \end{bmatrix}
  %= -i\tau_0\tau_3
  ,
  \quad
  %\\
  g_2
  %&
  \equiv\begin{bmatrix}-\sigma_y& 0\\ 0&\sigma_y\end{bmatrix}
  %= i\tau_1\tau_3
  ,
  \quad%\\
  g_3
  %&
  \equiv\begin{bmatrix}\sigma_z& 0\\ 0&-\sigma_z\end{bmatrix}
  %= -i\tau_3\tau_4
  ,
  \quad
  %\\
  g_4
  % &
  \equiv\begin{bmatrix}\sigma_z& 0\\ 0&\sigma_z\end{bmatrix}
  %= -i\tau_1\tau_2
  .\\
\end{gathered}
\end{equation}
It is also useful to express $g_{\alpha}$ in terms of $\tau_{\alpha}$:
$g_1 = -i\tau_0\tau_3$,
$g_2 = i\tau_1\tau_3$,
$g_3 = -i\tau_3\tau_4$,
and 
$g_4 = -i\tau_1\tau_2$.

%% for preparation, I prefer to use Bib TeX
%%\bibliography{holonomy,misc}

\begin{thebibliography}{21}
\bibitem{endnote:CEPL}
  We remark on the difference from the previous 
  report~\cite{Cheon:EPL-85-20001} on the notations concerning 
  to the analysis of the kicked spin-$\frac{3}{2}$.
  The spherical variables 
  $\gamma$, $\eta$, $\xi$ and $\zeta$ in Ref~\cite{Cheon:EPL-85-20001}
  should be read as $\theta$, $\frac{\pi}{2}-\eta$, $\chi$ and $\varphi$, 
  respectively.
  %%
  The set of normalized eigenvectors
  $\ket{v_{mn}}$ ($m, n = 0,1$) in Eq.~(36) of
  Ref.~\cite{Cheon:EPL-85-20001} and 
  $\ket{\xi_{\pm}}$ and $\ket{\TR \xi_{\pm}}$ are related by a unitary
  transformation:
  $\ket{v_{00}} 
  = \ket{\xi_+}\cos\frac{\eta}{2}+ \ket{\TR\xi_+}\sin\frac{\eta}{2}$,
  $\ket{v_{01}} 
  = \ket{\xi_+}\left(-\sin\frac{\eta}{2}\right)
  + \ket{\TR\xi_+}\cos\frac{\eta}{2}$,
  $\ket{v_{10}} 
  = \ket{\xi_-}\cos\frac{\eta}{2}
  + \ket{\TR\xi_-}\left(-\sin\frac{\eta}{2}\right)$,
  and
  $\ket{v_{11}} 
  = \ket{\xi_-}\sin\frac{\eta}{2}
  + \ket{\TR\xi_-}\cos\frac{\eta}{2}$.
  Accordingly the choices of basis vectors have difference.
  This modifies the expressions of the gauge connections and holonomy 
  matrices.
\end{thebibliography}

%% for submission, we need to embed the .bbl file here

%% end of .bbl

\end{document}